%% file: main.tex
\documentclass[pdftex,twocolumn,epjc3]{svjour3}          
\usepackage{todonotes}
\usepackage[backend=bibtex, style = ieee]{biblatex}
\usepackage{svg}
\usepackage{adjustbox}
\usepackage{siunitx}  
\usepackage[modulo,pagewise]{lineno}

\RequirePackage[T1]{fontenc}

\smartqed  

\RequirePackage{graphicx}
\RequirePackage{mathptmx}      
\RequirePackage{flushend}
\RequirePackage[colorlinks,citecolor=blue,urlcolor=blue,linkcolor=blue]{hyperref}

\def\makeheadbox{\relax}

\usepackage{amsmath}
\usepackage{amssymb}
\usepackage{listings}
\usepackage[super]{nth}

\journalname{Eur. Phys. J. C}

\begin{document}

\title{IceCube - Neutrinos in Deep Ice}

\subtitle{The Top 3 Solutions from the Public Kaggle Competition}

\author{Habib Bukhari, Dipam Chakraborty, Philipp Eller\thanksref{e1,addr1,addr2}, Takuya Ito, Maxim V. Shugaev\thanksref{addr3}, Rasmus Ørsøe\thanksref{e2,addr1,addr2}
}

\thankstext{e1}{e-mail: philipp.eller@tum.de}
\thankstext{e2}{e-mail: rasmus.orsoe@tum.de}

\institute{Technical University of Munich, TUM School of Natural Sciences, Physics Department, 85748 Garching, Germany\label{addr1}
          \and
Technical University of Munich, Munich Data Science Institute, 85748 Garching, Germany
\label{addr2} \and
University of Virginia, Department of Materials Science and Engineering, Charlottesville, VA 22904-4745, USA \label{addr3}
}

\date{}

\maketitle

\setlength\linenumbersep{5pt}
\renewcommand\linenumberfont{\normalfont\tiny\sffamily\color{black}}

\begin{abstract}
During the public Kaggle competition ``IceCube -- Neutrinos in Deep Ice'', thousands of reconstruction algorithms were created and submitted, aiming to estimate the direction of neutrino events recorded by the IceCube detector. Here we describe in detail the three ultimate best, award-winning solutions. The data handling, architecture, and training process of each of these machine learning models is laid out, followed up by an in-depth comparison of the performance on the kaggle datatset. We show that on cascade events in IceCube above 10\,TeV, the best kaggle solution is able to achieve an angular resolution of better than 5~degrees, and for tracks correspondingly better than 0.5~degrees. These performance measures compare favourably to the current state-of-the-art in the field.
\end{abstract}

\section{Introduction}
\label{sec:intro}
\input{sections/introduction_2}

\section{Competition Details \& Dataset}
\label{sec:competition}
\input{sections/the_dataset}

\section{Winning Solutions}
In this section, the technical details behind each winning solution is described. The details include core model architectures, choices in data preprocessing, standardization, training techniques and ensembling methods.
\label{sec:solutions}

\subsection{\nth{1} Place Solution}
\label{sec:solution1}
\input{sections/solution_1}

\subsection{\nth{2} Place Solution}
\label{sec:solution2}
\input{sections/solution_2}

\subsection{\nth{3} Place Solution}
\label{sec:solution3}
\input{sections/solution_3}

\section{Comparison of Solutions}
\input{sections/comparison}

\section{Conclusions}
\input{sections/conclusion}

\begin{acknowledgements}

This project was supported by the IceCube collaboration, kaggle.com and the Kaggle competition research grants program, the Munich Data Science Institute (MDSI), the Deutsche Forschungsgemeinschaft (DFG, German Research Foundation) under Germany's Excellence Strategy – EXC-2094 – 390783311, the Sonderforschungsbereich (Collaborative Research Center) SFB1258 ‘Neutrinos and Dark Matter in Astro- and Particle Physics’, and the PUNCH4-NFDI consortium fund “NFDI 39/1”. 

\end{acknowledgements}

\noindent \small{\textbf{Author Contributions} T.I. is the author of the \nth{1} place solution and detailed Sec.~\ref{sec:solution1}, H.B. and M.S. are the authors of the \nth{2} place solution and detailed Sec.~\ref{sec:solution2}, D.C. is the author of the \nth{3} place solution and detailed Sec.~\ref{sec:solution3}. R.\O. is the author of the DynEdge baseline. P.E. and R.\O. contributed all remaining text and edited the manuscript. R.\O. generated the model comparisons including the handling of all necessary data. P.E. is the main organizer of the original kaggle competition. All authors reviewed and discussed the full manuscript.
}

\printbibliography
\end{document}

%% file: sections/introduction_2.tex



The IceCube Neutrino Observatory~\cite{DetectorPaper} consists of a detector installed deep within the antarctic glacier and spans a cubic kilometer of ice. Its mission is to probe the properties of fundamental particles and the astrophysics of these particles. The main subject of study, so-called neutrinos, are the most abundant matter particle in the universe. They are nearly massless and do not carry an electric charge, making them particularly difficult to detect. An important step in analysing the data collected by the detector is to estimate the direction the neutrinos came from based on the measurements of the faint traces of Cherenkov radiation resulting from neutrino interactions in the ice.
This direction information is needed, for example, to unveil violent astrophysical neutrino sources \cite{IceCube:2014stg, IceCube:2018cha}, or to study neutrino properties \cite{icecube_osc, IceCubeCollaboration:2023wtb, IceCube:2021rpz}

\subsection{Reconstruction in IceCube}
Reconstruction of events is the process of turning the detector read-out data into high-level, physical quantities (such as the neutrino direction in our case), which is a parameter inference problem \cite{ml4astro}.
Traditional direction reconstruction algorithms in IceCube range from fast line fits \cite{linefit} to increasingly sophisticated maximum likelihood estimators (MLEs). A key component to MLE techniques is the event reconstruction likelihood itself that descrubes the scattering and absorption of photons in the South Pole ice, which is considered intractable and therefore requires approximation. Simpler techniques rely on parameterized distributions as approximations \cite{AMANDA:2003vtt}, or simplify the likelihood by describing only the direction reconstruction by removing pulses originating from scattered light \cite{santa}. Such algorithms are often used as first-guesses and fast reconstruction of real-time alerts \cite{icecube_real_time}.  More accurate but slower approaches make use of the full reconstruction likelihood, as described in \cite{icecube_waveform}. For instance Ref.~\cite{icecube_retro} is based on sophisticated photon ray tracing to approximate the full reconstruction likelihood, can reconstruct all event topologies, but is limited to the GeV energy range. 
Beyond the GeV scale, similar approaches are followed that typically target a specific type of event \cite{IceCube:2021rpz, IceCube:2021oqo}. Such methods have been widely used, most recently in finding evidence for neutrino emissions from the NGC1068 Seyfert galaxy~\cite{icecube_ngc1068}.

In recent years, new reconstruction techniques relying on neural networks (NNs) have emerged. In theory, NNs are able to approximate arbitrary functions, and once trained, they can reconstruct neutrino events many orders of magnitude faster than traditional methods. Techniques based on NNs are therefore both fast and flexible, as they can replace likelihoods, and can generalize to the entire energy range of IceCube. However, it has been an ongoing effort to identify model architectures that provide equal or superior reconstruction performance compared to the current traditional methods. This search is further complicated by the nature of neutrino telescope data; geometric time series is a data type that falls in-between established machine learning paradigms, making an a priori identification of ideal model architecture non-trivial.
Some early attempts used convolutional neural networks (CNNs) to reconstruct energy and direction of high-energy cascade events \cite{icecube_dnn}. A similar method was used for the energy reconstruction of neutrinos from NGC 1068 \cite{IceCube:2022der}. Also for the GeV energy range, an adaptation of CNNs have been used to a variety of reconstruction tasks \cite{icecube_flercnn_proceeding}. Graph Neural Networks (GNNs) have been shown to further improve the reconstruction accuracy for low energy events \cite{icecube_dynedge}, and have also been adapted for novel tasks such as pulse cleaning \cite{icecube_pulsecleaning}, but struggle to outperform traditional reconstruction methods beyond the GeV energy range.
Hybrid approaches combine MLE techniques with deep learning, where the likelihood is either fully or partly approximated by neural networks \cite{icecube_freedom, icecube_event_generator}. Such a technique was recently used to find evidence of neutrino emissions from the galactic plane \cite{icecube_galactic_plane}.

\subsection{``Neutrinos in Deep Ice'' kaggle competition}
Kaggle is an online platform where companies and institutions can present data science problems to the general public through competitions. In these competitions, members of the public can compete in teams to develop algorithms that perform best on a well-defined problem specified by the competition. 
We created the Kaggle competition ``IceCube -- Neutrinos in Deep Ice'' \cite{icecube-neutrinos-in-deep-ice}, where the participants were tasked with developing direction reconstruction algorithms for various IceCube neutrino events. Given a detector response $x$ which was induced by a neutrino with direction vector $r_{truth}$, the algorithms had to produce an estimate $r_{recon.}$ of the true neutrino direction. The scoring metric used to evaluate the quality of the algorithms was the mean opening angle between $r_{truth}$ and $r_{recon.}$ computed over a large set of events. 

During the three month competition period from January \nth{19} to April \nth{19} 2023, a total of 6460 people entered the competition and 901 submitted at least one valid solution.  At the end of the competition, the participants were distributed across 812 competing teams and a grand total of 11,206 solutions had been submitted~\cite{icecube_kaggle_proceeding}, of which the top three, prize-winning solutions are presented in this article.

%% file: sections/the_dataset.tex
\begin{figure}[h!]
    \begin{center}
    \includegraphics[width=0.4\textwidth]{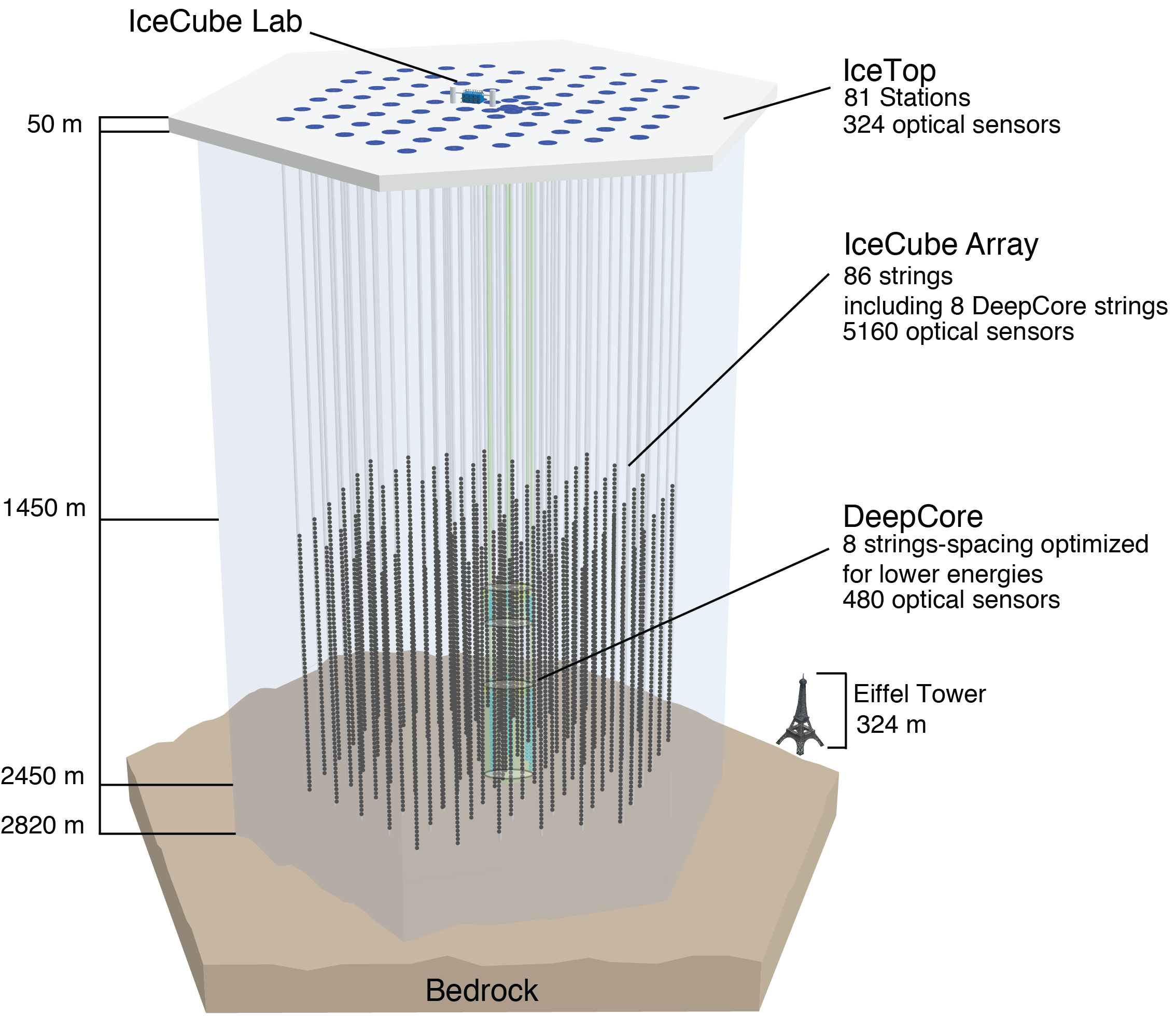}
    \caption{Illustration showing the IceCube detector with its 86 strings vertically deployed in ice, instrumenting the depth between 1450\,m and 2450\,m. The 78 strings of the main array are arranged in a nearly hexagonal pattern, while the remaining 8 strings have a denser sensor spacing forming ``DeepCore'' and are highlighted in green. (Image courtesy of the IceCube collaboration.)}
    \label{fig:icecube_array}
    \end{center}
\end{figure}

The IceCube detector \cite{DetectorPaper} consists of 5160 Digital Optical Modules (DOMs) \cite{icecube_daq} distributed on 86 strings at depths from 1450\,m to 2450\,m, see Fig.~\ref{fig:icecube_array}. The main array of the detector consists of 78 strings arranged in a near-hexagonal pattern each carrying 60 DOMs with a vertical separation of 17\,m and an average horizontal distance between neighbouring strings of 125\,m. Each DOM holds a 10" Photomultiplier Tube (PMT) directed towards the center of the Earth. At a depth of around 2000\,m, a layer of optical impurities lies embedded in the ice. The layer is mostly comprised of mineral dust, referred to as "the dust layer" and effects the scattering and absorption of light \cite{dust_layer}. An additional 8 strings have been installed around the center string of the main array. On these strings, the DOMs are distributed differently than in the main array: Above the dust layer, at around 1750\,m to 1850\,m, 10 DOMs comprise the so-called "veto cap", a cluster of DOMs used to identify electrically charged particles entering the detector from above. Below the dust layer, from around 2100\,m to 2450\,m, a part of the ice where the optical transparency is highest, a second cluster of DOMs have been installed. This second cluster, named ``DeepCore'' \cite{deepcore}, uses DOMs with higher quantum efficiency compared to the main array and the modules have a vertical spacing of just 7 meters, making it the detector volume with the highest density of DOMs. 

When neutrinos interact in the ice, charged particles are produced that emit Cherenkov radiation as they transverse the ice. The PMTs in IceCube DOMs can detect these photons, and the amount of signal detected for a neutrino interaction may range from a few to more than $10^5$ photons. The number of detected photons, however, is many orders of magnitudes lower than those emitted, and this signal is interspersed with noise from primarily radioactive decays in the glass housing of the DOMs. 

\subsection{Neutrino Events in IceCube}
When photo-electrons produce a sufficiently large voltage at the PMT anode, a digitization process is triggered where the PMT waveforms are read out either partially or fully. If at least one neighboring DOM on the same string also records a signal within 1\,$\mu$s, the hard local coincidence (HLC) condition is satisfied and the full waveform is read out. If this condition is not met, only minimal information around the peak voltage is read out \cite{DetectorPaper}.
The digitized waveforms are subject to an unfolding process \cite{icecube_waveform} that estimates photon arrival times and the charge of individual photo electrons---each of which is referred to as a so-called ``pulse''. 
Based on the HLC hits recorded by the DOMs, event triggers set certain criteria for reading out all signal recorded by DOMs for further processing. Different event trigger definitions are used in the IceCube online systems \cite{icecube_trigger}. 

Neutrino events in IceCube come mainly in two broad categories that have distinct geometric shapes. ``Tracks'': sufficiently energetic charged-current (CC) muon neutrinos (and roughly 17\% of $\nu_{\tau}$ interactions) produce a track signature. These events produce a muon that can travel long distances in the ice while emitting Cherenkov radiation, effectively producing a track-like signature.``Cascades'': Other neutrino interactions that are not described by the track class. These interactions produce electromagnetic and hadronic particle showers in which the energy tends to be deposited over relatively small distances, effectively producing a localized deposit of light. 

\begin{figure}[h!]
\centering
   \includegraphics[width=0.45\textwidth]{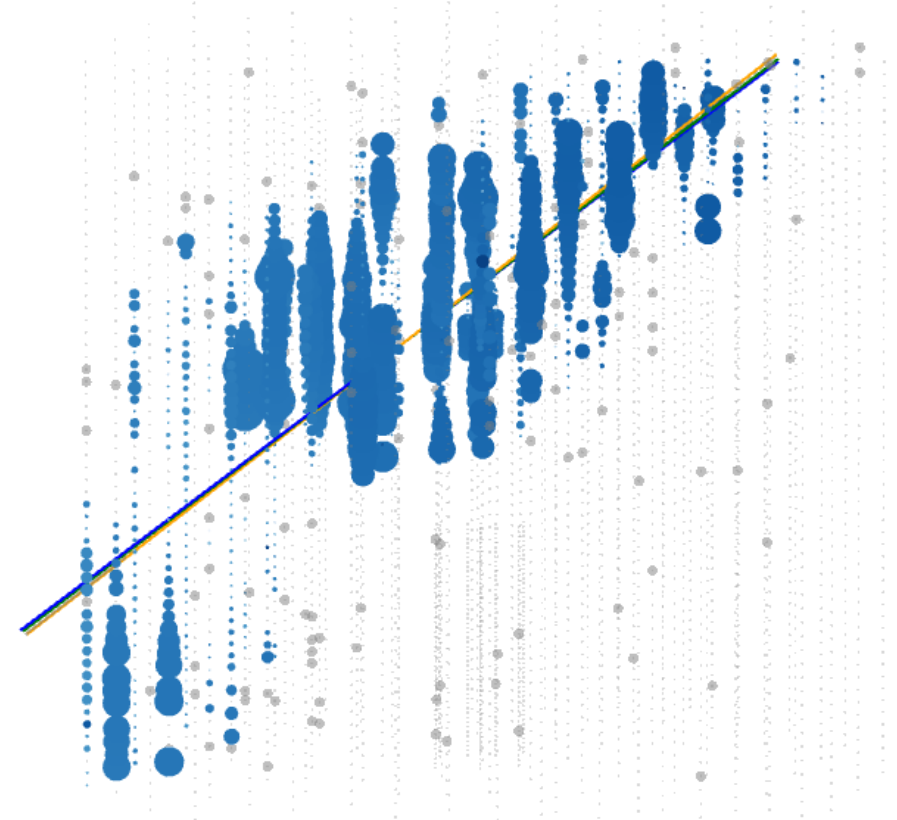}
   \includegraphics[width=0.45\textwidth]{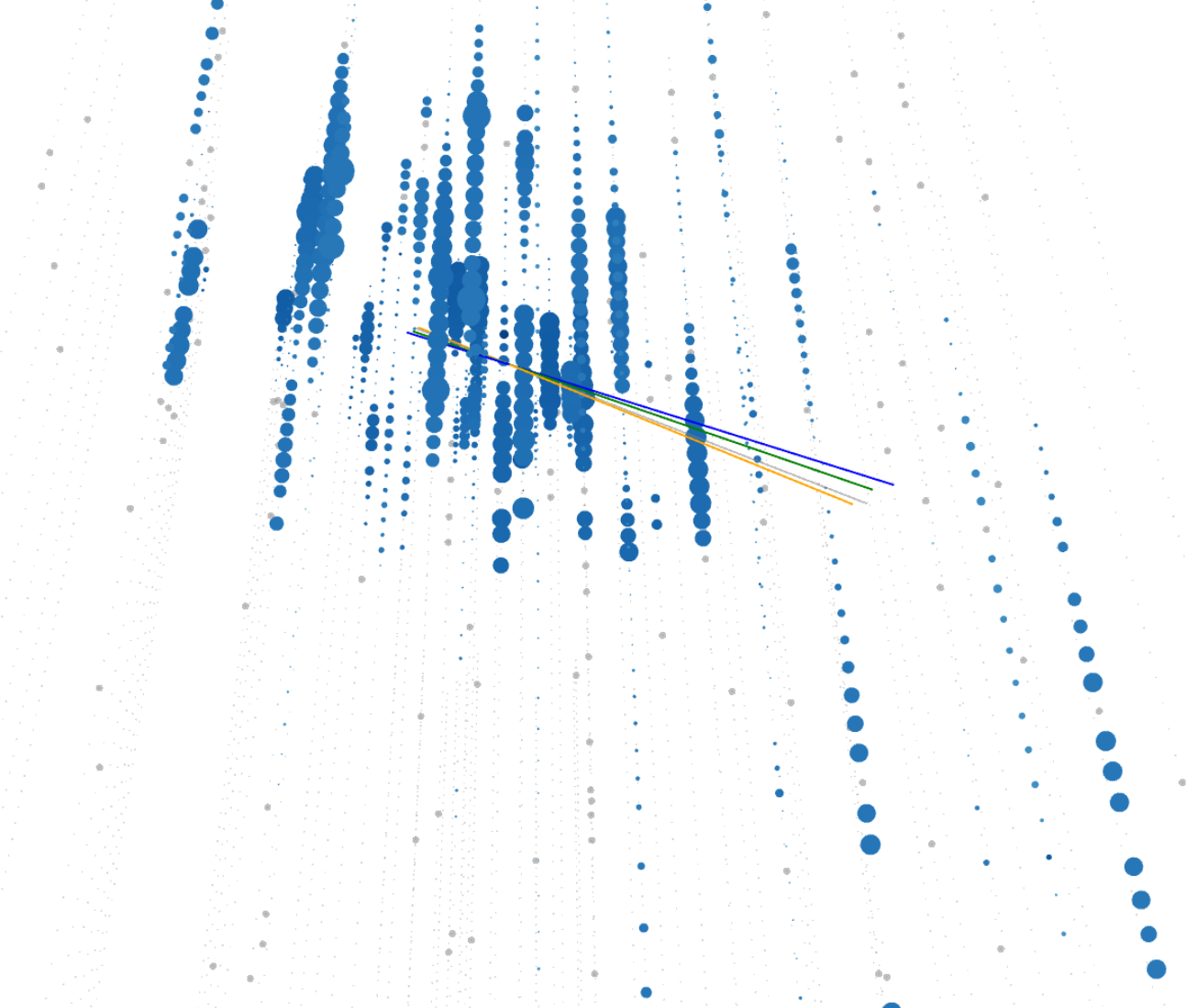}
   \label{fig:event_display}

\caption{A 22.4 PeV $\nu_\mu$\,CC event from the kaggle dataset where the neutrino interacted outside the detector volume, allowing the muon to travel from above and through the detector with zenith and azimuth angles of 52 and 189 degrees, respectively. Non-auxiliary (i.e. HLC) pulses are shown in blue. Solid grey points indicate auxiliary (i.e. non-HLC) pulses, likely induced by noise. Not shown is the time information of the pulses. Neutrino direction reconstructions are shown with coloured lines, and the grey line depicts the true neutrino direction. The reconstructed directions were (53, 190), (52, 188), (53,189) for the \nth{1}, \nth{2} and \nth{3} place solution respectively. \textbf{Top:} A side view of the event -- the muon passes through the entire detector. \textbf{Bottom:} Close-up of the end of the track highlighting the differences in predictions.}
\end{figure}

An illustration of a simulated track event is shown in Figure \ref{fig:event_display}. In this event, the neutrino interacted outside the detector volume, producing a muon that enters the detector from above. Because the neutrino had 22 PeV of total energy, the resulting muon is able to travel through most of the detector and a large number of DOMs measure Cherenkov radiation from the interaction. The true direction and reconstructed directions from each winning solution is added for comparison. 

The participants of the kaggle competition were provided with nearly 140 million simulated neutrino events together with the true directions.  These events span the energy range from 100\,GeV to 100\,PeV, and contain all flavours and interaction types, and hence both tracks and cascades. These events were grouped into 660 random sub-samples, so-called "data batches", each containing the detector response from 200.000 events.  Each detector response $x$ is a [$\text{n}_\text{pulses}$, 6]-dimensional array, where $\text{n}_\text{pulses}$ is the number of observed pulses in the trigger window, which can range from a few to several hundreds of thousands. For each pulse, the in-ice position of the PMT, arrival time,  the associated charge and an auxiliary flag is provided. These six features are the input to the reconstruction algorithms and are shown in Table~\ref{tab:features}. 
\begin{table}[h]
\centering
\caption{Input data for Kaggle Competition: Neutrinos in Deep Ice}
          \label{tab:features}
    \begin{tabular}{lll}
    \hline
    Feature & Description & Unit  \\ 
    \hline  
     $(x,y,z)$ & Position of DOMs in IceCube coordinates & m\\
     $t$ & Pulse time relative to trigger time & ns \\
     $q$ & Charge of a pulse & P.E. \\
     $Aux.$ & If 0, the pulse is in HLC & - \\
     \hline
    \end{tabular}
\end{table}
If the auxiliary flag is 1 it indicates that the specific pulse did not meet HLC criteria, and is thus more likely to originate from noise. Each event is simulated using a unique set of nuisance parameters that represents the systematic uncertainties of the detector. This includes assumed scattering and absorption of light, and lets the collection of events represent a wide range of detection scenarios  \cite{icecube_snowstorm}. In addition, the detector responses have not undergone any pulse cleaning, a procedure that attempts to remove noise-induced pulses.

\subsection{A Competition Baseline With DynEdge}
To provide the participants with a baseline to compare against, we decided to train a GNN from \textsc{GraphNeT} \cite{graphnet}, an open-source ML library for neutrino telescopes, on parts of the competition data. We submitted the predictions from the GNN to the leaderboard for comparison, and shared the trained model and technical material that allowed participants to fully tinker with every aspect of this method. Many solutions, including the winning solutions documented in this paper, took inspiration from the techniques in our baseline submission and produced their own versions aimed specifically for this competition. In this section we elaborate on some of those techniques.

\subsubsection{DynEdge}

The specific GNN called ``DynEdge'' \cite{icecube_dynedge}, is a flexible algorithm capable of reconstructing and classifying many different physics tasks on both a per-pulse and per-event level. DynEdge is a convolutional graph neural network which represents neutrino event as point cloud graphs. A graph is a collection of nodes $n$ and edges $e$. In the competition baseline, we represented individual pulses of Cherenkov radiation as nodes, and edges are initially drawn to each node's $k$ Nearest Neighbours ($k$NN) based on the Euclidean distance between the in-ice PMTs that measured the pulse. The data associated with each node is the pulse information shown in Table \ref{tab:features}. The graphs are then convolved  by an EdgeConv \cite{edgeConv} layer, which updates the feature vector of the $i$'th node by adding the pair-wise differences between $i$'th node and each of its $k$ neighbours. The layer is defined by
\begin{equation}
\mathbf{x}^{\prime}_i = \sum^{j=k}_{j \in \mathcal{N}(i)} h_{\mathbf{\Theta}}(\mathbf{x}_i, \mathbf{x}_j - \mathbf{x}_i)
\label{edgeconv_equation}
\end{equation}
where $\mathbf{x}_i$ represents the un-convolved feature vector of the $i$'th pulse,  $\mathbf{x}_j$ the feature vector of the $j$'th neighbour of the $i$'th pulse and $h_{\mathbf{\Theta}}$ is a learned function applied to all neighbourhoods. The convolved values of the $i$'th node is represented by $\mathbf{x}^{\prime}_i$. DynEdge applies multiple of these layers in series, and between each layer the neighbourhoods are re-calculated based on each node's position in latent space, effectively letting the GNN learn the optimal edges for the given task. The most recent use of DynEdge in IceCube is in a study of IceCube Upgrade's expected sensitivity to atmospheric neutrino oscillations, where it was used to remove noise pulses, classify event topologies and reconstruction \cite{icecube_pulsecleaning}. 

\subsubsection{Von Mises-Fisher Loss}

In our baseline submission we used the von Mises-Fisher (vMF) distribution for 3D vectors as a loss function \cite{icecube_dynedge}. By taking the natural logarithm, one obtains
\begin{equation}
\text{loss} = -\kappa \cdot  \cos(\Delta \theta) + \ln(C_3(\kappa))
\label{vmf_loss}
\end{equation}
where $\Delta \theta$ is the opening angle between the true and reconstructed 3D direction vector, and $\kappa$ is a measure of uncertainty, analogues to $\frac{1}{\sigma}$ from normal distributions. The quantity $C_3(\kappa)$ represents the normalization constant of the vMF distribution, which requires numerical approximation for 3D vectors. The vMFs distribution for 3D vectors represents a 2-sphere embedded in $\mathbb{R}^3$ and is conceptually close to conventional choices in loss functions such as $1 - \cos(\Delta \theta)$ but has the added benefit of the uncertainty estimation through $\kappa$, which is estimated alongside the direction by DynEdge. 

For the baseline, we trained DynEdge according to the procedure documented in \cite{icecube_dynedge} using the 3D vMF loss on 7.8\% of the competition data without further optimization. It achieved a score of around 1.018, which is calculated as the mean opening angle taken over the evaluation dataset. Using the full data set, the score drops to around 0.985.

%% file: sections/solution_1.tex
We designed a simple, lightweight model combining the EdgeConv~\cite{edgeConv} and the transformer architecture, allowing us to leverage the strengths of both architectures.
By creating six variations of this model and ensembling them, we achieved a private leaderboard score of 0.960.

\subsubsection{Preprocessing}
Transformers can achieve a high accuracy even without extensive feature extraction when provided with adequate training data. 
Given this, our approach emphasized retaining the inherent characteristics of the data, applying minimal preprocessing.

The input features are as defined in Table~\ref{tab:features}.
Due to computational constraints, the sequence for each event was truncated to a maximum of 6000 pulses for inference, and only events with a maximum of 200 to 500 (depends on model) pulses or less were used for training.
Each input feature is scaled as follows.
\begin{align}
& x', y', z' = \frac{x}{500}, \frac{y}{500}, \frac{z}{500} \\
& t' = \frac{t - 10^4}{115} \\
& q' = \frac{\log_{10}(q)}{3}
\end{align}

In addition to these elementary features, node homophily ratio \cite{homophily} of $(x, y, z, t)$ is extracted as global statistics.
These global statistics features are concatenated with the output from the global pooling layer shown in Figure \ref{fig:1st_place_model}.
These preprocessing steps were similar to those of DynEdge \cite{icecube_dynedge} 

\subsubsection{Base Model Architecture}
\begin{figure}[h]
    \begin{center}
    \includegraphics[width=0.5\textwidth]{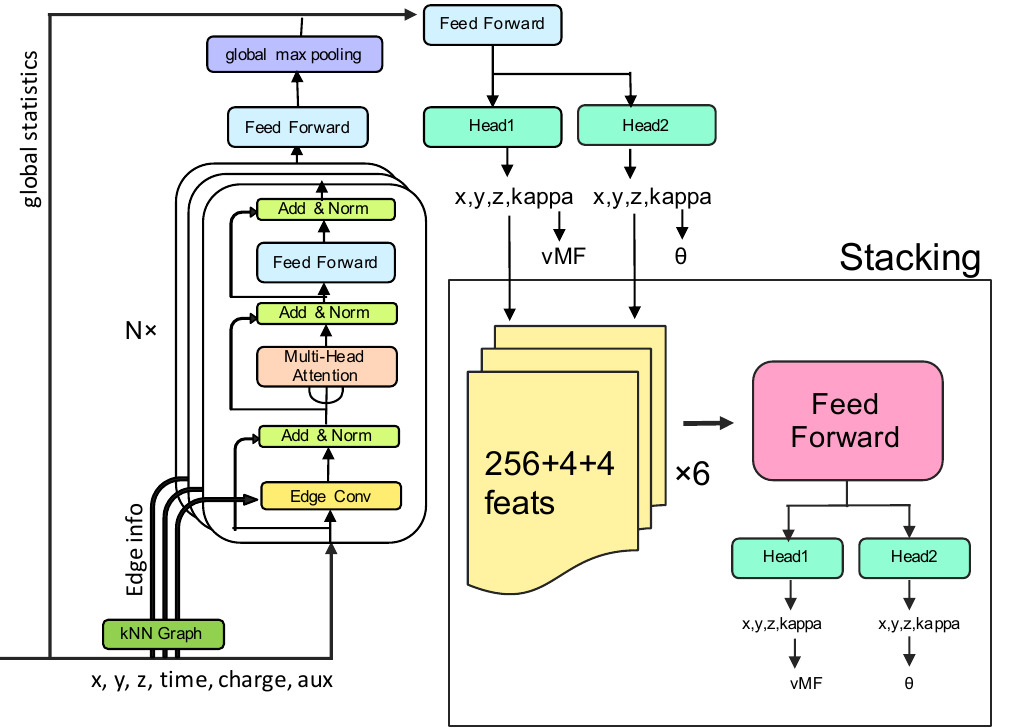}
    \caption{This illustrates the overall architecture of the \nth{1} place model. The left part represents the core structure of our method, combining EdgeConv and a Transformer layer, while the right part depicts the stacking process.}
    \label{fig:1st_place_model}
    \end{center}
\end{figure}

An overview of our model is shown in Figure~\ref{fig:1st_place_model}.
The base model uses multiple blocks of EdgeConv followed by a transformer.

\subsubsection{EdgeConv}
In the original EdgeConv layer implementation, the feature vector $\mathbf{x}_i$ associated with DOM$_i$ is updated based on the difference $\mathbf{x}_j-\mathbf{x}_i$, where \( \mathbf{x}_j \) represents the feature vector of the $k$-nearest neighbor DOM$_j$ to DOM$_i$, as seen in Eq. \ref{edgeconv_equation}. This scheme works well for features like x, y, z, and time. But for features like charge and the auxiliary flag, its absolute values are important. Therefore, we updated \( \mathbf{x}_i \) based on both the difference \( \mathbf{x}_j-\mathbf{x}_i \) and \( \mathbf{x}_j \) itself.
\begin{equation}
\mathbf{x}^{\prime}_i = \sum^{j=k}_{j \in \mathcal{N}(i)} h_{\mathbf{\Theta}}(\mathbf{x}_i, \mathbf{x}_j - \mathbf{x}_i, \mathbf{x}_j)
\end{equation}

\subsubsection{Edge Selection}

In the original EdgeConv implementation, edges are calculated in each layer dynamically by $k$-Nearest Neighbors ($k$NN). However, this edge selection scheme is not differentiable in itself and therefore does not have gradients.
This would still work well for the segmentation task in the original paper, as points in the same segment are trained to be close in the latent space. However, the situation is different in this task, where it would not make sense to dynamically select edges. Therefore, the edges used in EdgeConv are calculated from the input features only once in our model.

As illustrated in Figure \ref{fig:1st_place_model}, our method first calculates the edges of a neutrino event. The event graph containing node features $x$ and edges $e$ is then passed through EdgeConv which uses the edges to convolve the node features. These latent, convolved node features $\mathbf{x}^{\prime}$ are then added together with the original, unconvolved node features $x$ in a skip-connection, and the result is normalized. The normalized quantity is passed to a transformer with multiple attention heads, and this output is also subject to skip-connection addition and normalization. The output is given to a MLP block with a final addition and normalization skip connection.  This combination of EdgeConv and a Transformer layer forms the backbone of our method and may be repeated in serial $n$ times, as denoted in the diagram. The Number of layers $n$ is a hyperparameter which has been subject to tuning, and our final method utilizes multiple instances of this base model with different choices in $n$ in an ensemble. The output of this series of EdgeConv + Transformer blocks is fed through an MLP block and is subject to global max pooling, producing a latent column vector for each event. This column vector is concatenated with \textit{global statistics}, i.e. a set of engineered features that describe the entire event. This column vector is given to a final MLP with two prediction heads. 

\subsubsection{Loss Function}
While the von Mises Fisher Loss shown in Eq. \ref{vmf_loss} serves as a reliable and consistent loss function, it represents \( \theta \) using cosine values. However, the metric of this competition is the angle \( \theta \) itself.
To minimize \( \theta \) itself, we defined the loss function as follows:
\begin{equation}
\text{Our Loss} = -\theta - \kappa \cos(\theta) + ln(C_3(\kappa))
\end{equation}
This simple modification resulted in a 0.005 decrease in opening angle compared to von Mises-Fisher Loss.

\subsubsection{Ensemble Members}
We made six models for ensemble, and their configuration can be seen in  Table \ref{table:model_configurations}.
The dimensions of both EdgeConv and transformer layers are set 256 and the number of attention heads for the transformer is set to 8 for all ensemble members.
The number of EdgeConv+Transformer block in each model are either 3 or 4. The variables used to calculate the distance used for edge selection by $k$NN are either the 3d position $(x, y, z)$ or the 4d input $(x, y, z, t)$ and 6 edges are extracted for each node. Maximum sequence lengths for the transformer layers range between 200 and 500 pulses pr. event for training, but is set to 6000 pulses per event for inference.
A maximum of about 30 epochs was used as we found training beyond this point gave increasingly diminishing returns.
\begin{table}[!h]
\caption{Overview of ensemble members and their configurations. $n_l$: Number of EdgeConv+Transformer block layers in serial connection. $k$: The positional information used for computation of edges, 3d denoting the three spatial dimensions and 4d denoting the inclusion of time. $b_s$: Batch size used for training. $s_l$: Maximum input sequence length used for training. $n_e$ Number of epochs. $O$: Origin of model, "-" indicates training from scratch.  $S_{valid}$.: Local validation score.  $S_{public}$ and $S_{private.}$ are the public and private leaderboard scores, respectively.} 
\begin{adjustbox}{width=\columnwidth}
\begin{tabular}{|l|r|r|r|r|r|l|r|r|r|}
\hline
\textbf{ID} & \textbf{$l$} & \textbf{$k$} & \textbf{$b_s$} & \textbf{$s_l$} & \textbf{$n_e$} & \textbf{$S_{valid.}$} & \textbf{$S_{public}$} & \textbf{$S_{private}$} & \textbf{O}\\
\hline
M1 & 4 & 3d & 1000 & 200 & 30 & 0.967 & 0.964 & 0.964 & -  \\
M2 & 4 & 3d & 1000 & 200 & 2  & 0.967 & 0.964 & 0.963 & M1 \\
M3 & 3 & 3d & 500  & 300 & 30 & 0.969 & 0.965 & 0.965 & - \\
M4 & 3 & 4d & 1000 & 250 & 30 & 0.971 & 0.968 & 0.968 & - \\
M5 & 4 & 3d & 1000 & 200 & 12 & 0.966 & 0.963 & 0.963 & M2 \\
M6 & 4 & 4d & 500  & 500 & 30 & 0.968 & 0.964 & 0.964 & - \\
\hline
\end{tabular}
\end{adjustbox}
\label{table:model_configurations}
\end{table}

These models were selected from experiments conducted during the limited competition period and were not designed to achieve the best result.
In particular, model dimension and depth of EdgeConv+Transformer block is very small, model parameters are only 6M for the largest model, and we believe that a larger model would give better performance.

\subsubsection{Ensemble Method: Stacking}
Our approach to ensembling incorporates a stacking strategy, where the predictions from the models listed in Table \ref{table:model_configurations} are used as input for a final model. The six ensemble members have been trained with varying configurations, each capturing different nuances of the data. By using these as embedding extractors, the stacking model can benefit from the diverse representations and deliver a more comprehensive prediction.

The stacking model is a three-layered Multi-Layer Perceptron (MLP) with dimensions 512. It takes as input the prediction of $x$, $y$, $z$, \( \kappa \) and the last 256-dimensional hidden layer from each of the ensemble members. The detailed architecture of this process is illustrated in Figure \ref{fig:1st_place_model}. The same loss functions used in base models are used for this stacking model. Through stacking, a further decrease in average opening angle by 0.003 degrees was achieved.

\subsubsection{Training Procedure}

The Adam optimizer \cite{KingmaBa:2014adam} was used for all models.
For models M1, M3, M4, and M6, which were trained from scratch, the learning rate was set to  \(10^{-3}\) for the first 15 epochs, 
and then it was linearly decayed from \(10^{-3}\) to \(10^{-6}\) for the subsequent 15 epochs. 
For the fine-tuned models M2 and M5, in the initial half of their training, they had a learning rate of \(10^{-4}\), and during the latter half, their learning rate was linearly decayed from \(10^{-4}\) to \(10^{-7}\).
To effectively train our models on this voluminous dataset, we employed several optimization techniques. Specifically, by incorporating Mixed-Precision Training \cite{mixed_precision} and Sequence Bucketing, we managed to reduce the GPU memory consumption to about a third, while achieving over three times faster computational speed. This approach allowed us to handle the data very efficiently while preserving the accuracy of our models.

\subsubsection{Mixed-Precision Training}
Mixed-precision training is a technique where input data and model weights are cast from the high-precision 32-bit to 16-bit floating point precision during training. This is known to both decrease memory usage and speed up computations.
When we first used mixed-precision, it made the training unstable. 
However, by appropriately placing the batch normalization at specific positions as shown in Fig.~\ref{fig:1st_place_model}, we achieved stable progression in the training process. 

\subsubsection{Sequence Bucketing}
\begin{figure}[h]
    \begin{center}
    \includegraphics[width=0.5\textwidth]{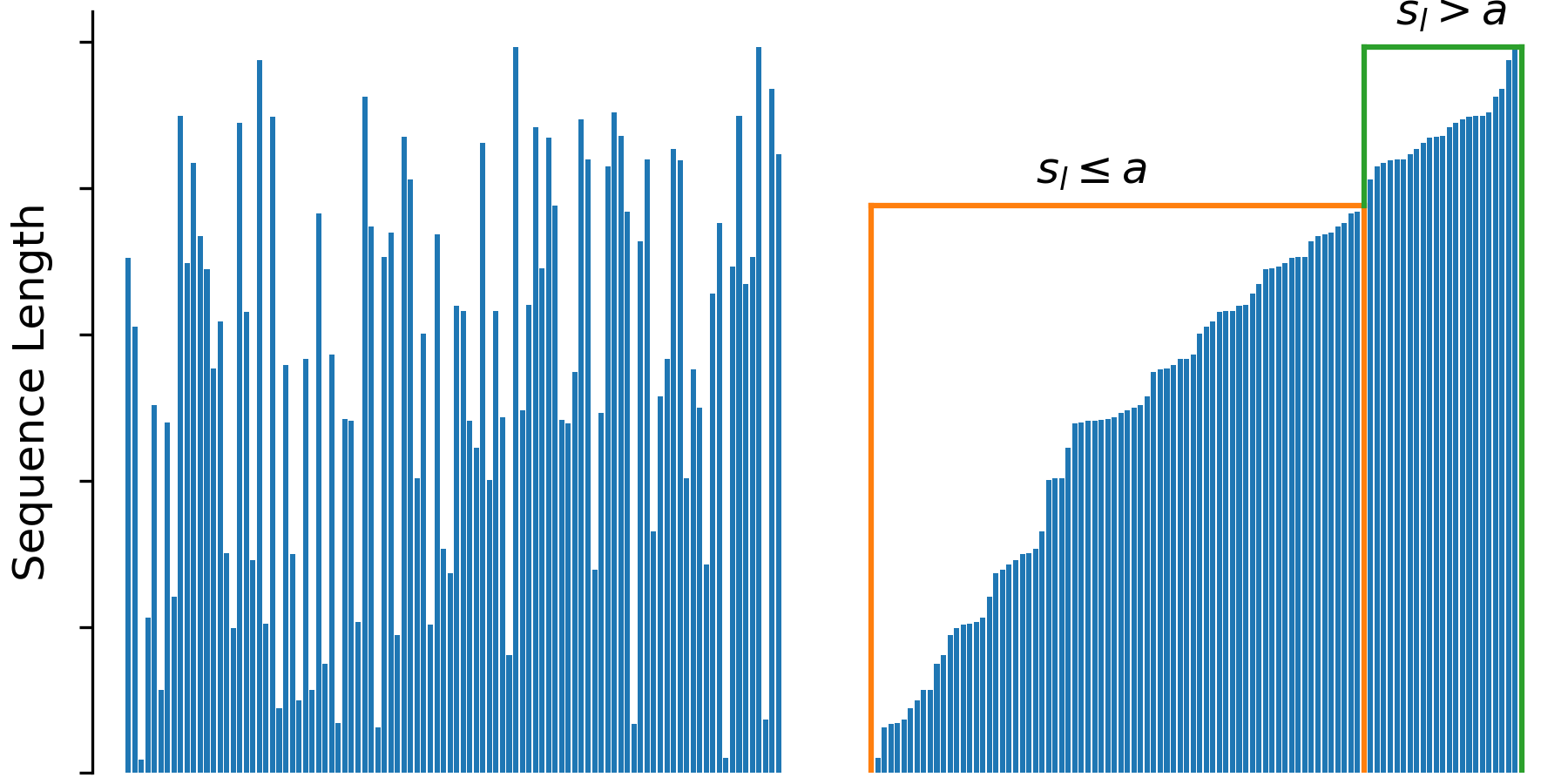}
    \caption{Illustration of sequence bucketing for sequence data with varying lengths. The y-axis denotes the sequence lengths, whereas the x-axis denotes the order of appearance in a batch of training data. \textbf{Left}: Unordered set of sequences in a batch. \textbf{Right}: The set of sequences are sorted according to their lengths and sliced into two "buckets". $S_l$ denotes the sequence length and $a$ represents a threshold choice.}
    \label{fig:1st_place_Sequence_Bucketing}
    \end{center}
\end{figure}

In transformer models, the self-attention mechanism inherently has a quadratic computational complexity with respect to sequence length. 
As sequence lengths increase, this leads to significant computational and memory challenges. Sequence bucketing addresses this issue by grouping sequences of similar lengths together in "buckets" \cite{sequence_bucketing}. As illustrated in Fig.~\ref{fig:1st_place_Sequence_Bucketing}, this approach minimizes the need for excessive padding, optimizes GPU memory use, and reduces the number of unnecessary computations. Consequently, using sequence bucketing can enhance the computational efficiency of training transformer models without compromising on performance. In our training, we chose the threshold $a$  in Figure \ref{fig:1st_place_Sequence_Bucketing} such that the samples within a batch are split in a \( 8:2 \) ratio. This ratio was determined experimentally to minimize the GPU memory usage.

In addition, some ensemble members were not trained from scratch, but further trained from other ensemble members. M2 was fine-tuned from M1 and M5 was fine-tuned from M2. Out of the 660 batches of competition data, the last batch was used for local validation of ensemble members during training. The batch size was set as large as possible to fit in GPU memory, ranging from 500 to 1000. A maximum of about 30 epochs was used for all ensemble members as we found training beyond this point gave increasingly diminishing returns, however some models trained for fewer epochs as seen in Table \ref{table:model_configurations}.

%% file: sections/solution_2.tex
In our pursuit to better predict the direction of neutrino particles based on data from the IceCube Neutrino Observatory, we have critically assessed the constraints of conventional GNNs. These include their predominantly local function and unexpectedly slow computational processing in comparison to transformer models. In particular, in our tests, 1.4M parameters and 4 layer DynEdge reference model requires the same computational budget for training as 16 block and 7.4M parameters transformer \textit{T} model, with the latter providing a significant improvement of the mean opening angle. Therefore, our proposed solution hinges on transformer models, which we see as evolved GNNs that operate on a fully connected graph, dynamically estimating edge weights via attention mechanisms \cite{attention_all_you_need}. In addition to transformers, a Fourier encoder is central to our method, which embeds the continuous input variables, shown in Table \ref{tab:features}, into a multidimensional space used in the model. This addition significantly improved the quality of our direction reconstructions. Also, our method relies on a DynEdge-inspired feature extractor and a custom implementation of the Minkowski space-time line element as an attention bias, which refined our method further. Our code is available as open source \cite{2nd_place_github}.

\subsubsection{Preprocessing}
This section details the techniques used for data standardization, feature engineering, and data subselection.
  
\subsubsection{Standardization Techniques}
Each of the detector coordinates was divided by \(500\). The time was shifted by \num{1e4} towards the beginning of the event and then was divided by \num{3e4}. For the charge, we took base-10 logarithm and divided the result by 3. This standardization is similar to the reference DynEdge model \cite{icecube_dynedge}. For the length of the event, we took the base-10 logarithm of the total number of pulses. The Minkowski space-time interval used the same normalization as the detector coordinates. The ice properties data was taken from \cite{icecube_ice}. The depths were adjusted by subtracting 1950\,m and subsequently normalized by dividing by 500\,m. The scattering and absorption lengths were standardized using the \texttt{RobustScaler} from the scikit-learn library \cite{scikit-learn}, ensuring the robustness of the transformation to outliers.

\subsubsection{Feature Engineering}

The input to the model consists of a sequence of pulses described by the variables in Table~\ref{tab:features}. The base-10 logarithm of the total number of pulses is included as an event-level feature to provide the model with information on the event length in case of under-sampling of long events, as described in the next subsection.  

We found it crucial to process the continuous variables, \textit{e.g.} time and charge, into a representation suitable for transformers using the Fourier encoding method. Fourier encoding is a technique from signal processing and is frequently used for language models to describe the position of a word in a sentence \cite{attention_all_you_need}. This encoding method, however, generalizes beyond natural language processing and, when applied to a continuous signal, can be viewed as soft digitization of the input into a set of Fourier sine/cosine modes with \begin{math}10000^{2j/d}\end{math} frequencies, where \textit{d} is the embedding width, and integer \textit{j} is changing from 0 to \begin{math}d/2 -1\end{math} \cite{attention_all_you_need}. 

In our setup, the continuous input variables undergo the Fourier encoding method, while the discrete auxiliary variable is transformed using a learnable embedding. By multiplying the normalized time and position by 4096 and charge by 1024, we attained sufficient digitization resolution. Specifically, this resolution is dictated by the highest considered Fourier frequency, which is equal to 1. A change of the normalized input continuous variable by \begin{math}2\pi\end{math} results in the identical value in this lowest bit of the digitization. The model may capture lower variations of the input due to the continuous nature of the Fourier encoding, but the signal change may be too weak for the model to be treated effectively. For example, with the above-described scaling, we increased the sensitivity of the embedding to small input variations and can achieve \num{3e4} ns\begin{math}/4096\times2\pi\end{math} = 1.2 ns digitization resolution for time (\num{3e4} ns comes from the time normalization). This input up-scaling is critical, and the use of the Fourier encoding method with sufficiently large scaling coefficients significantly improved the model performance in our experiments.

In addition to Fourier encoding, in part of our experimental setups we have incorporated an optional DynEdge-inspired encoder with several adaptations, which provides an additional set of latent features as input to our base model.  

Lastly, to combat the noise induced pulses, we further introduced a relative space-time interval bias based on the Minkowski metric given by \begin{math}ds^2=c^2 dt^2-dx^2-dy^2-dz^2\end{math}. The metric is used to compute the line element between all pulses in a given neutrino event and encodes their causality. We define the element as \begin{math}ds=sign(ds^2) \cdot \sqrt{|ds^2|}\end{math} clipped at $(-4,4)$ and to achieve higher digitization resolution, we divide $ds$ by 1024 before passing $ds$ through the Fourier encoding. This engineered feature is added to the first transformer blocks in our base architecture as a relative attention bias \cite{sa_with_rel_pos_repres}, which provided a noticeable improvement in the performance.

In addition to the feature engineering above, we explored the impact of including ice transparency and absorption as additional features. Despite these extended features offering an alternate representation for the $z$-coordinate, because of their $z$-dependence, no noticeable improvement to the mean opening angle was found when including the optical properties of the ice.

\subsubsection{Sub-selections and Filtering}

The IceCube challenge considers relatively short events with 62 median and 163.4 average number of pulses, and only 1.4\% of events are longer than 768 pulses. Therefore, we trained our models with the maximum sequence length of 192 and used 512 lengths in validation and 768 in the final submission. In scenarios where events exceeded the pre-determined maximum sequence length, a tiered selection mechanism was employed. The primary preference was given to HLC pulses and in cases where the number of available HLC pulses fell below the pre-determined sequence length, non-HLC pulses were sampled. Other sampling techniques based on charge and arrival time were investigated but did not improve upon the choice described above. Consideration of longer sequences may improve the model performance on cascade events with a large number of pulses.

\subsubsection{Base Model Architecture}

\begin{figure*}
    \centering
    \includegraphics[width=\textwidth]{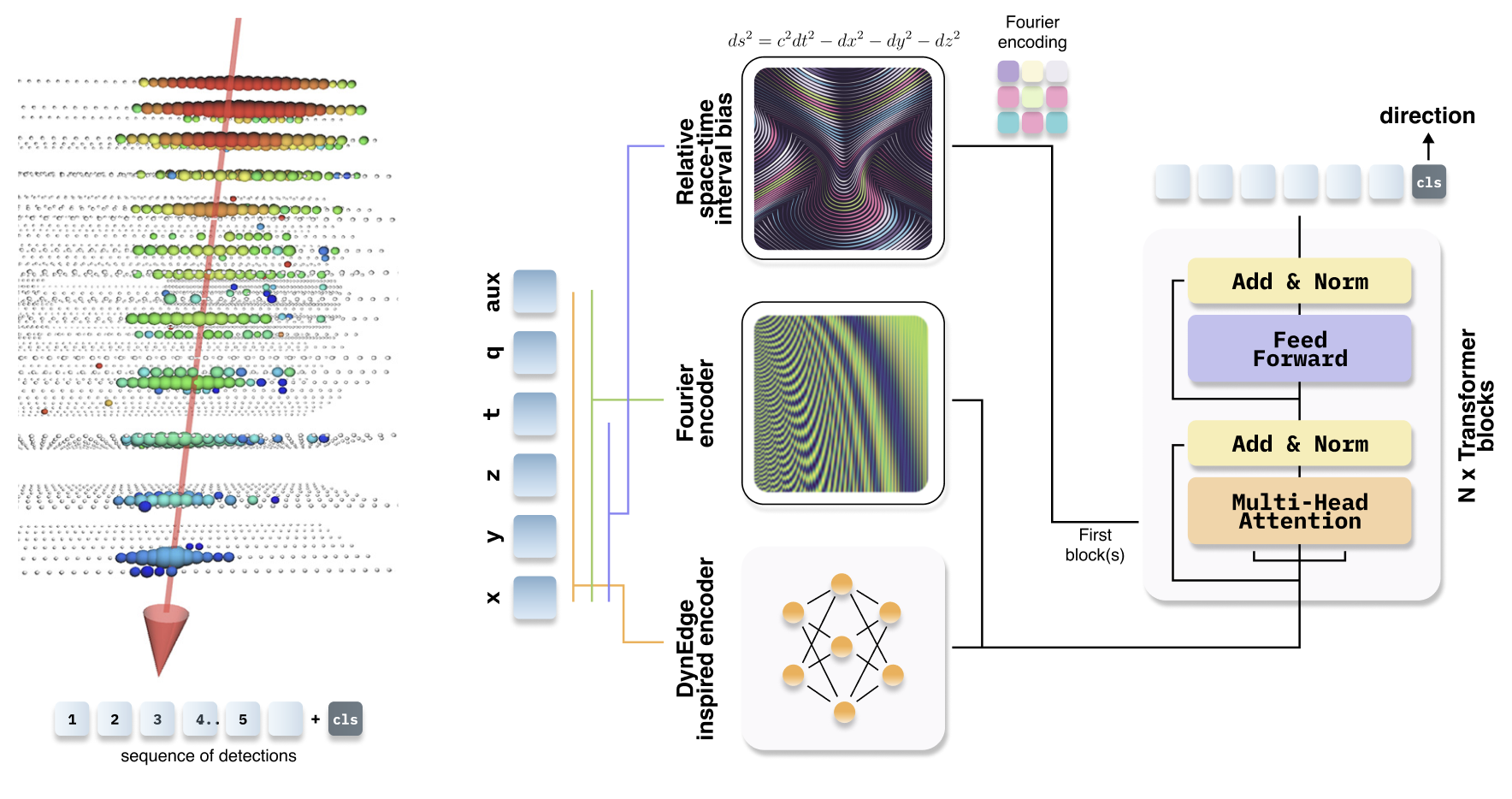}
    \caption{Illustration of the \nth{2} place transformer-based setup for interpreting events as sequences of pulses. The flow begins with feeding pulses to the Fourier encoder to process continuous input variables. In parallel, there is an option to use the DynEdge-inspired encoder for additional feature extraction. The outputs from these stages are then concatenated. The first 4 transformer blocks employ relative space-time interval bias for clustering pulses based on their causality. In later blocks, the input sequence is expanded with a \textit{cls} token, which is projected at the end into a 3-dimensional vector with the neutrino direction.}
    \label{fig:model_2nd_place}
\end{figure*}

Central to our method is a transformer model that interprets each event as a sequence of pulses, and the base model diagram can be seen in Figure \ref{fig:model_2nd_place}. We use transformer blocks with learnable shortcuts, similar to BEiT \cite{beitv2}, and the first 4 transformer blocks are modified according to \cite{sa_with_rel_pos_repres} in order to incorporate the relative space-time as attention bias. The continuous input variables are processed with the Fourier encoder and the optional DynEdge-inspired encoder to create features, which are then concatenated and fed to the transformer. In the DynEdge encoder, we switched all ReLU activations to GELU. Unlike the standard implementation, we opted not to employ pooling operations and instead utilized the latent features from the encoder. In the first 4 blocks of the transformer, relative space-time intervals are given the Fourier-encoded as attention bias. 

The first 4 transformer blocks are followed by 12 regular transformer blocks with learnable shortcuts. The input sequence to these blocks is expanded with a \textit{cls} token \cite{devlin2019bert}, which is a special token often used in transformer architectures, primarily to represent the entire input sequence for tasks like classification. In the context of our model, the \textit{cls} token is targeted to aggregate the information about the neutrino direction when the data propagates through the transformer blocks. After the last transformer blocks, this \textit{cls} token is projected into a 3-dimensional vector, which characterizes the predicted direction of the neutrino and the model's confidence in the prediction (the vector length). 

The details of the models, considered in our experiments, are summarized in Table \ref{tab:2nd_model_specs}. Our model sizes are referred to according to ViT \cite{touvron2021training} notation: \textit{T} for tiny (192), \textit{S} for small (384), and \textit{B} for base (768). The computational cost of the smallest considered \textit{T} model is similar to the competition DynEdge baseline.

\begin{table}
\centering
\caption{Model Specifications Summary. In the "Depth Conf." column, the depths are represented in the order: Relative Bias Blocks, Dynamic Edge Blocks, and Transformer Blocks. Each entry denotes the number of blocks (or layers for DynEdge-inspired encoder) in each respective component. A dash "-" indicates that a component is not present in the model.}
\label{tab:2nd_model_specs}
\begin{tabular}{|c|c|c|c|c|}
\hline
Model & Dims & Head Size & Depth Conf. & Parameters \\
\hline
T & 192 & 32 & 4/-/12 & 7.57M \\
S & 384 & 32 & 4/-/12 & 29.3M \\
B & 768 & 32 or 64 & 4/-/12 & 115.6M \\
S+DynEdge & 384 & 32 & 4/4/8  & 23.3M \\
B+DynEdge & 768 & 32 or 48 & 4/4/12  & 116.5M \\
\hline
\end{tabular}
\end{table}

\subsubsection{Ensemble Members}
Several experiments were carried out to determine the effect of the model size, pooling mechanism, head size, and use of the Fourier and DynEdge-inspired encoders on the model performance and derive the optimal sets of hyperparameters for the base model architecture. Our experimental efforts are summarized in the table below for a series of key configurations. In our evaluations, the Cross-Validation (CV) score was gauged at \(L=512\) maximum sequence length (denoted as CV512). The last five data batches (655 through 659) served as our validation set. Columns $S_{public}$ and $S_{private}$ refer to the evaluation on the test data of the IceCube challenge for the public and private leaderboard, respectively. The models included in our final ensemble are highlighted in bold.

\begin{table}
\centering
\caption{Performance Breakdown of Model Configurations Across Benchmarks. \textit{d} in the Model column refers to the head size. Models with an asterisk (*) were trained on 2×A6000 hardware, a distinct choice from the RTX4090 that permits a significantly larger actual batch size. Models used in our final ensemble are highlighted in bold, and the $W_e$ column represents the weight used for each ensemble member in the final linear combination of predictions.}
\label{tab:2nd_model_performance}
\begin{tabular}{|c|c|c|c|c|}
\hline
Model & $W_e$ & CV512 & $S_{public}$ & $S_{private}$ \\
\hline
T d32 & - & 0.9704 & 0.9693 & 0.9698 \\
T d32 no Fourier & - & 0.9900 & 0.9883 & 0.9871 \\
T d32 avr pool& - & 0.9705 & 0.9693 & 0.9692 \\
S d32 & - & 0.9671 & 0.9654 & 0.9659 \\
\textbf{B d32}& 0.0825 & 0.9642 & 0.9623 & 0.9632 \\
\textbf{B d64}& 0.1535 & 0.9645 & 0.9635 & 0.9629 \\
\textbf{B+DynEdge d48} & 0.1937 & 0.9643 & 0.9624 & 0.9627 \\
\textbf{*S+DynEdge d32}&  0.2360 & 0.9639 & 0.9620 & 0.9628 \\
\textbf{*B+DynEdge d32}& 0.3343 & 0.9633 & 0.9609 & 0.9621 \\
\hline
\end{tabular}
\end{table}

\textit{T d32} is our baseline model used for the optimization of the training pipeline, which provides a significant boost over the DynEdge reference model. To disentangle the effect of the Fourier encoder on the performance, we have performed an experiment \textit{T d32 no Fourier} using a simple projection of the continuous input variables into the transformer dimension. This experiment results in a significantly lower mean opening angle suggesting that it is crucial to convert the continuous input variables into a form suitable for transformers. The next experiment \textit{T d32 avr pool} uses a simple masked average pooling over all tokens in the sequence instead of \textit{cls} token based pooling. This experiment results in nearly identical performance to our \textit{T d32} baseline, and we have chosen \textit{cls} token setup for all further experiments. \textit{S d32} and \textit{B d32} experiments highlight the effect of the model size on the performance, suggesting an approximately 0.003 decrease in the mean opening angle with each increase of the model from \textit{T} to \textit{S} to \textit{B} size. This upscaling, however, carries the trade-off of a four-fold increase in model parameters and the computational cost. In addition, we have noticed the onset of overfitting in training B models, and the size of the dataset provided for the IceCube challenge may be insufficient for training B and larger models. \textit{B d64} experiment illustrates the effect of the head size on the model performance, suggesting that a smaller head size of 32 is preferable to 64, typically used in comparable size language and vision transformer models \cite{devlin2019bert, dosovitskiy2021image}. The next series of experiments (\textit{B+DynEdge}, \textit{*S+DynEdge}, and \textit{*B+DynEdge}) are targeted at the study of the effect of the DynEdge-inspired encoder. These experiments exhibit rather contradictive results and may be affected by the use of different hardware, which permits a larger actual batch size. Overall we see a small improvement or no improvement from using DynEdge-inspired encoder. Our best final submission to the competition was an ensemble of 5 experiments highlighted in bold in Table \ref{tab:2nd_model_performance} because their combination provided the best validation score. 

\subsubsection{Ensembling Method}
Our best final submission achieved a score of 0.9594 at the public and 0.9602 at the private test sets (0.9610 CV512). We considered a simple weighted average of the 3-d vectors predicted by the models with weights listed in Table \ref{tab:2nd_model_performance}, which are fitted to maximize the validation score. The use of the weighted average is partially dictated by the property of von Mises-Fisher Loss (applied in training) that enforces the model to correlate the vector length with the confidence \cite{vml}. Therefore, a weighted average of the model predictions automatically biases the final prediction toward the most confident direction. Ensembling only offered a marginal improvement over our best single model result, which stands at 0.9608 public and 0.9618 leaderboard (0.9627 CV512). We also considered stacking with fitting a second-level few-layer neural network model combining predictions of the first-level models (Table \ref{tab:2nd_model_performance}) and additional features, \textit{e.g.} sequence length, first detection time, and the total charge. However, these experiments did not bring any improvement over the simple weighted average baseline.

\subsubsection{Training Procedure}
For the effective management and utilization of the dataset provided by the organizers, several data loading techniques were employed during training, each summarized below.

\begin{itemize}
\item \textbf{Data Chunk Caching}: The provided dataset \cite{icecube-neutrinos-in-deep-ice} is too large to be fully stored in RAM on small workstations. On the other hand, loading each data chunk and grouping the pulses based on the event takes a substantial time and makes it prohibitively expensive to perform random access to the dataset during training. To mitigate the computational overhead from data loading and preprocessing, a caching mechanism was employed, \textit{i.e.} recently used data chunks are stored in RAM, while the older chunks are removed. This strategy enables fast random access to data from recent chunks and at the same time drastically reduces the RAM requirement compared to storing the entire dataset.

\item \textbf{Chunk-Based Random Sampling}: 
The limitation of the plant chunk caching, however, is that under random sampling there is only little chance that the same chunk is sampled multiple times within a short time, and the computational overhead on data loading is still significant. To enable effective cache utilization, we have implemented a strategy that selects a random data chunk first and then samples randomly all events within this chunk before going to the next one, \textit{i.e.} each data chunk is loaded only once per epoch while the sampling is randomized. This strategy provides a good balance between random access and utilization of computational resources.

\item \textbf{Sequence Bucketing}: All sequences in the input batch (not to be confused with "data batches" used above to refer to individual files in the dataset) must be padded to the longest sequence length in the batch before feeding to a transformer model. Meanwhile, the processing of these padding tokens brings an additional computational overhead. For random batch sampling (the default option in deep learning libraries, \textit{e.g.} Pytorch \cite{pytorch}) each batch has a substantial chance of having at least one sequence significantly exceeding the median sequence length of the dataset. Processing of padding tokens in a such batch may take several times more computational budget than processing of actual sequence tokens. For example, the competition dataset has a median sequence length of 62 while the maximum sequence length in our training is set to 192. To minimize this computational overhead we adopted a length-matching batch sampling strategy that constructs batches of events with approximately the same length (buckets of the length multiple of 16). This sequence bucketing has enabled training with the maximum sequence length of 192 and inference at lengths up to 768 without a substantial increase in computational time. Using a sequence length of 768 during inference yielded a noticeable performance improvement when compared to the 192 length.
\end{itemize}

Data chunks 655 to 659 were reserved for local validation, while the remaining data chunks were used in training. In all experiments, the AdamW optimizer \cite{adamw} with a weight decay of 0.05 was utilized. We employed a cosine annealing scheduler with a warm-up, adjusting the learning rate according to a cosine function over the course of training. The scheduler was reset at each epoch. The effective batch size was maintained at 4096 using the gradient accumulation \cite{gradientaccu} technique to navigate hardware limitations.

The use of the competition metric (the mean opening angle) as the loss function is preferable because it enables the direct minimization of the objective value. However, model ensembling may favor the incorporation of the model confidence into the prediction, which is provided in training with von Mises-Fisher loss \cite{vml} as the length of the predicted vector. In our training in the first 2-3 epochs, we used the von Mises-Fisher loss. In the remaining epochs, the model is fine-tuned with $l = \mathrm{opening\ angle} + 0.05 \cdot vMF$ loss to optimize the competition metric (we still retained a factor of 0.05 from the von Mises-Fisher loss to facilitate ensembling).

The training was carried out for 4-5 epochs depending on the loss plateau, but the most notable performance improvement occurred within the first epochs. At the initial epoch, \textit{T} and \textit{S} models had the maximum learning rate of \(5 \times 10^{-4}\), whereas the \textit{B} models utilized a rate of \(1 \times 10^{-4}\). In the following epochs, the maximum learning rate was reduced from \(2 \times 10^{-5}\) to \(0.5 \times 10^{-5}\). Mixed precision is used to accelerate the training. From the second epoch onward, the incorporation of Stochastic Weight Averaging (SWA) \cite{swa}, a method that averages model weights across multiple training steps for better generalization, was helpful in improving the model performance.

Training time varies according to the model configuration. Specifically, the \textit{T} model with 7.57M parameters takes about 10 hours per epoch, whereas the \textit{B+DynEdge} model with 116.5M parameters required 56 hours per epoch on an \textsc{Nvidia RTX4090}.

%% file: sections/solution_3.tex
Recent literature that applies machine learning to IceCube data often use Graph Neural Networks (GNNs). DynEdge presents neutrino events as point cloud graphs where edges are dynamically learned in small neighbourhoods, thus treating reconstruction of neutrino events as a dominantly geometric learning problem. We posit that treating IceCube data as a sequence learning problem is the optimal choice. Following this, we make use of the transformer architecture in our solution. Emperically we found transformers to have higher utilization of GPUs, and run faster than GNNs both during training and inference, despite doing more computations. We also find that transformers scale well with the amount of data, closely following known scaling laws that guide the training of large language models \cite{hoffmann2022empirical}. We model angle prediction of IceCube data as both a classification and regression task in our solution. We found each to have their own strengths and weaknesses. Finally, we combine both predictions using scikit-learn's \textsc{HistGradientBoostingClassifier} \cite{scikit-learn}, a method that uses an ensemble of decision trees. The training\footnote{https://github.com/dipamc/kaggle-icecube-neutrinos} and inference\footnote{https://www.kaggle.com/code/dipamc77/3rd-place-attention-xgboost-ensembler} code for our solution is open-sourced. Figure \ref{fig:solution_3} depicts the high level architecture of the solution.

\subsubsection{Preprocessing}
We treat each event as a sequence of pulses sorted by their arrival time from early to late. We represent the neutrino events as sequences with lengths up to 3072. If an event has more than 3072 pulses, we prioritize HLC pulses. For events with more than 3072 HLC pulses, we randomly sample 3072 HLC pulses. For each event, we set the time of first pulse of the sampled sequence to 0.  Each element in the sequence is a vector of pre-processed data about each pulse.

\subsubsection{Standardisation Techniques}
The $x$, $y$, and $z$ coordinates are each divided by 500. Relative time values are divided by 30000. We take the logarithm of the charge values and divide it by 3. These choices are similar to the standardization done for DynEdge.

\subsubsection{Feature Engineering}
We have engineered features specific for each ensemble member in our final method. 
First, for the transformer model we add quantum efficiency of the PMTs, and optical properties of the ice near each sensor taken from \cite{icecube_daq}. While we use the features mentioned in the final models, we note the performance without these features in early experiments done with small transformers only gave a small benefit than that without the features.
\begin{table}[h]
\centering
\caption{Features used in the gradient boosting model. Along with statistics of the pulse information, and the actual predicted angles from both the classifier and regressor, we included some hand engineered features. Features marked with * indicate custom hand engineered features. The primary motivation for the features is to mitigate the bias of the zenith classifier. We categorized HLC pulses that are on vertically adjacent sensors and close in time as HLC pulse pairs. Then we compute the features based on this categorization, note that we ignore HLC pulses in this categorization if they do not have a corresponding paired pulse. From the regressor predictions, we compute kappa, which is the square root of the inverse norm of the predicted vector.}
\label{tab:3rd_boosting_features}
\begin{tabular}{p{0.1\linewidth} | p{0.6\linewidth}}
\hline
\textbf{Feature} & \textbf{Description} \\
\hline
 $z_{\text{stats}} $ & min, max, mean and  std z-coordinate  \\
 \hline
 $t_{\text{stats}} $ & min, max, mean and std of absolute values arrival time \\
 \hline
 $q_{\text{stats}}$ & std and sum of charge \\
 $P_{\text{hlc}} $  & Percent of total pulses being HLC  \\
 $S_{\text{avg.}} $ & average speed between pulses \\
 \hline
 $P_{pair}$ *   & Percentage of pulses that are HLC pulse pairs \\
 $\Delta_{12}$ *   & Difference of time and z-coordinate between first HLC pulse pair and second HLC pulse pair\\
  $Z_{1}$ *  & Z value of first HLC pulse pair \\
 \hline
 $\theta$ & Actual angles predicted by the classifier and regressor \\
$\kappa$ & kappa - Proxy for confidence of prediction\\
 $S_{A}$ & Softmax of azimuth classifier predictions \\
 $S_{Z}$  & Softmax of zenith classifier predictions \\
 $\Delta \theta$ & Angular distance between all direction predictions \\
\end{tabular}
\end{table}
Second, for the gradient boosting model, we added many features, as shown in \ref{tab:3rd_boosting_features}. Most of the features do not seem to affect the performance. Shap values indicate only few are important, namely $\kappa$, first pulse absolute time, and disagreement between predicted angles.
Third, we prepared the zenith and azimuthal angles as binned truth values for the angle classifiers. The angles are quantized into 128 bins and the spacing between azimuth bins is uniform. The spacing between zenith bins is uniform in cosine space.

\subsubsection{Base Model Architecture}

\begin{figure}[h!]
    \begin{center}
    \includegraphics[width=\columnwidth]{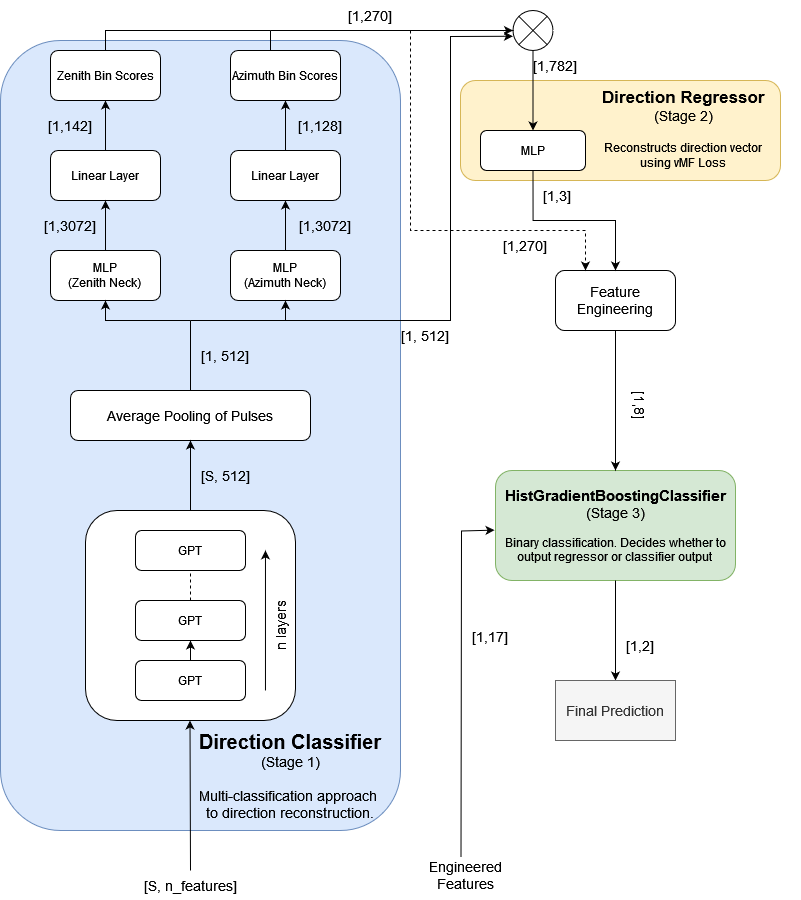}
    \caption{Model architecture of the \nth{3} place solution. We use a multi-staged architecture to produce the angle for each event. The main component is the transformer backbone, which is trained from scratch on preprocessed events. Each event is modelled as a sequence of pulses for which the transformer produces a sequence of embeddings. These embeddings are then average pooled and used to produce independent classification predictions of azimuth and zenith. The same embeddings are also used to do perform regression for the 3D vector corresponding to the angle. Finally, the regression and classifiation predictons are combined using a gradient boosting classifier, which takes the predictions as well as some hand engineered features as it inputs. The final solution uses three independent copies of this architecture, and their predictions are combined to produce the predictions we submitted to the competition.}
    \label{fig:solution_3}
    \end{center}
\end{figure}

We use a multi-stage approach to predict the angles for each event, as shown in Figure \ref{fig:solution_3}. The main component of our model is the transformer. We use standard GPT \footnote{https://github.com/karpathy/minGPT} \cite{radford2019language} layers as the main building block. We preprocess and optionally subsample each sequence to as the input to the transformer, which produces a sequence of latent vectors for each element in the sequence. This sequence is then average pooled to produce a single vector, that feeds into classification heads. The transformer predicts the classification of azimuth and zenith with separate classification heads. Subsequently, the classification outputs and averaged pooled vector are passed into the regression head, which predicts a 3D vector corresponding to the angle of the event. Finally, the predictions of both the regression and classification heads concatenated with a set of hand-engineered features, and passed to an Gradient boosting classifier, which is essentially an ensemble of decision trees.

For the transformers, each layer contains 8 attention heads, with an embedding size of 512 (64 per head). The transformer layers form the backbone of the model, producing an embedding for each pulse in the sequence. These embeddings are then averaged to make a single embedding per sequence. This embedded sequence is passed to an MLP with hidden dimension 12288 and output dimension 3072, which we call the neck. The neck for the classification head and regression head are separate. The neck embeddings are passed to the final layers for classification and regression.

\textcolor{red}{} 

The total number of trainable parameters of the 18 layer 
classification model is 72M. The transformer backbone has 57M parameters and the neck has 15M parameters.
The classification head is comprised of two independent fully connected layers of 128 values for azimuth and 142 (128 + 14) values for zenith. Note that extra 14 zenith values are due to the specific design of our loss function, which we describe in the training details subsection below.
The regression head is comprised of a single fully connected layer of 3 values, for predicting the 3D vector corresponding to the azimuth and zenith. Additional to the averaged embeddings of the sequence from the transformer backbone, the regression head gets the outputs of the prediction head as inputs, for a total of 782 ($=512 + 128 + 128 + 14$) inputs.

To combine the predictions of classification and regression heads, we use a gradient boosting model, specifically scikit-learn's \textsc{HistGradientBoostingClassifier} \cite{scikit-learn}. The boosting model receives hand engineered features built from the predictions and the event data, and does a binary classification of whether to use the regression or classification head for a given event. These engineered features are shown in Table \ref{tab:3rd_boosting_features}.


We noticed that a significant boost in performance is also obtained due to the feature of first pulse absolute time. Note that the transformers are trained only with relative time information, setting the first pulse time to 0. We did not experiment with absolute time values as inputs to the transformers, but results of other participants indicates that using absolute time gives a significant improvement. 

\subsubsection{Ensembling}


Our final solution was an ensemble of three copies of the same architecure, which is a full instance of the model shown in Fig.~\ref{fig:solution_3}. One of the models is a 15 layer transformer, and the other two are 18 layer transformers. The 18 layer transformers had weights taken from the same training run, where one was taken when 90 percent of the training was complete, and the other when the the training was fully completed, hence their predictions had high agreement. The final predictions of each model are combined using the \textsc{HistGradientBoostingClassifier}, note that this is separate from the gradient boosting classifier used for combining classification and regression heads.


Each of the three ensemble members independently produce the azimuth and zenith angles for each event. The predictions, along with all the hand engineered features of each event described in the feature engineering subsection above, are passed into the classifier, which only decides which of the three predictions to use for each event. We tried other ensemble methods such as averaging the predictions, but found the gradient boosting model to be most effective.

\subsubsection{Training Procedure}

The training of the 18 layer backbone transformer and classifier model was done on a single \textsc{Nvidia RTX4080} GPU over approximately 5 days. The 15 layer model takes about 10 percent less time to train than the 18 layer. For efficiency, we train the transformer and the classifier separately, which takes nearly all the training time. We also perform fine tuning of the trained model on longer sequences, which takes an additional 4 hours. The regression head is trained after the classification model training is complete, and the backbone is kept frozen, training the regression head takes 6 hours on the same hardware. The gradient boosting model takes only a few minutes to train. 
The training procedure can be broken into a few stages, outlined below.
\begin{itemize}
\item \textbf{Stage 1}: 
In the first stage, we train the transformer from scratch for 4 epochs on 650 of the 660 databatches provided by the organizers. We train for a maximum sequence length of 256, this was a practical choice due to the constraints of our hardware. 
We use the softmax cross-entropy loss with a custom smoothing operation for the classification loss. Unlike the standard softmax cross entropy setting, our classes are adjacent in angular space, and hence are not independent. To address this, we apply smoothing using a 1D Gaussian kernel for the local bins. The Gaussian kernel is of length 15 and sigma 3. For azimuth, since the space forms a closed loop, the smoothing operation also wraps around. For zenith, we pad the space with extra values to allow the guassian kernel to smooth beyond the boundary angles. At inference time, any predictions beyond the zenith boundary limits are clipped.
The learning rate scheduler, weight decay and optimizer play an important role in the final performance. For classification models, we tune these parameters on smaller models and scale down the learning rate for larger models. We use a maximum learning rate of \num{1e-4} and a weight decay of \num{1e-5}, with a OneCycle learning rate schedule, the final decayed learning rate is \num{2e-6}. We use AdamW optimizer for all the models. We keep a batch size of 384 for the sequence length of 256, and use gradient accumulation where needed, to fit out VRAM constraints. We also do gradient clipping with norm of 0.5 to stabilize training.

\item \textbf{Stage 1 Fine-tuning}: The classification models are fine-tuned for longer sequences with 20 data batches for 1 epoch. We increase the maximum sequence length of events from 256 to 3072 in this stage, and only train on the events that are higher than 256 in length. During inference, we only use the weights of the fine-tuned model for the sequences longer than 256.
Since the subset of data above 3072 is quite small, the total number of events that was used for fine-tuning is small, and the training is completed in 4 hours.

\item \textbf{Stage 2}: After completing the classifier training, we run inference on 100 data batches to save the outputs. Regression model is trained on these 100 data batches for 35 epochs. Since the regression head was trained with the backbone frozen, it is significantly faster to cache the outputs of the backbone and directly train the regression head. Caching outputs training the regression head takes 6 hours.
For the regression head, we use the von Mises-Fisher loss shown in Eq.~\ref{vmf_loss}. In our experiments, we found it to be unstable when training the full transformer backbone. Hence we opted to the freeze the backbone after training of the classification head, and only train the regression neck and head. The vMF also allows to use the kappa feature as a proxy for prediction confidence. Kappa is analogous to the inverse of the norm of the predicted vector.
The choice of hyperparameters didn't affect performance much for the regression models. We keep the same hyperparameters as the classification models, except the batch size which was increased to 1024.

\item \textbf{Stage 3}: In the final stage the gradient boosting classifier is trained with the outputs of both the regressor and classifier. We save predictions of 2 data batches from both models, and train the \textsc{HistGradientBoostingClassifier} with the labels set to the predictions of whichever model was closer to the ground truth for every event.
Our primary motivation to add this stage was the observation that the distribution of predictions for zenith and azimuth had unwanted structure for both the classifier and regressor. Through experimentation, we found that these predictions were independently not close to the ground truth, but when combined using the \textsc{HistGradientBoostingClassifier} in this stage, the performance was significantly improved.  
\end{itemize}






In adition, we employ several training optimizations to reduce the training time. We used Flash Attention v1 \cite{dao2022flashattention} during transformer training, which reduced the memory requirement and training time. The memory reduction allowed usage of longer sequence length for fine tuning. We train the transformer in float16 precision. Approximately halfway through the training, we switch to float32 precision for stability. We also used sequence bucketing by sorting the dataset by sequence length and packing mini-batches by sequence length, the longest sequence in a batch will determine the padding for the rest of the batch.



%% file: sections/comparison.tex
\begin{figure}[h!]
    \begin{center}
    \includegraphics[width=0.35\textwidth]{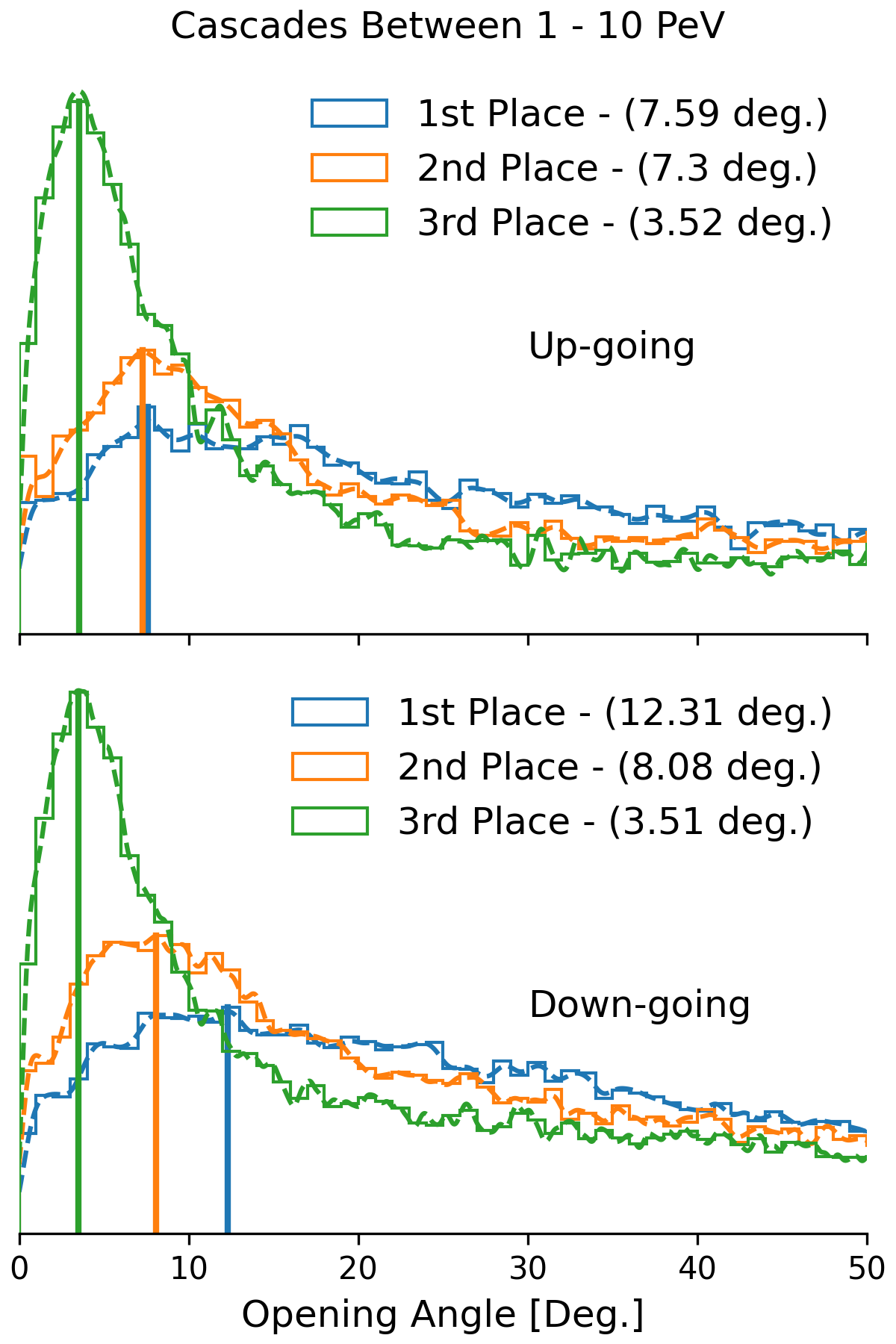}
    \caption{Example of  KDEs constructed for each solution on cascade events between 1 and 10 PeV. Estimated primary mode is denoted with the vertical lines. \textbf{Top:} Up-going cascade events. \textbf{Bottom:} Down-going cascade events.}
    \label{fig:kde_cascades}
    \end{center}
\end{figure}
\begin{figure}[h!]
    \begin{center}
    \includegraphics[width=0.35\textwidth]{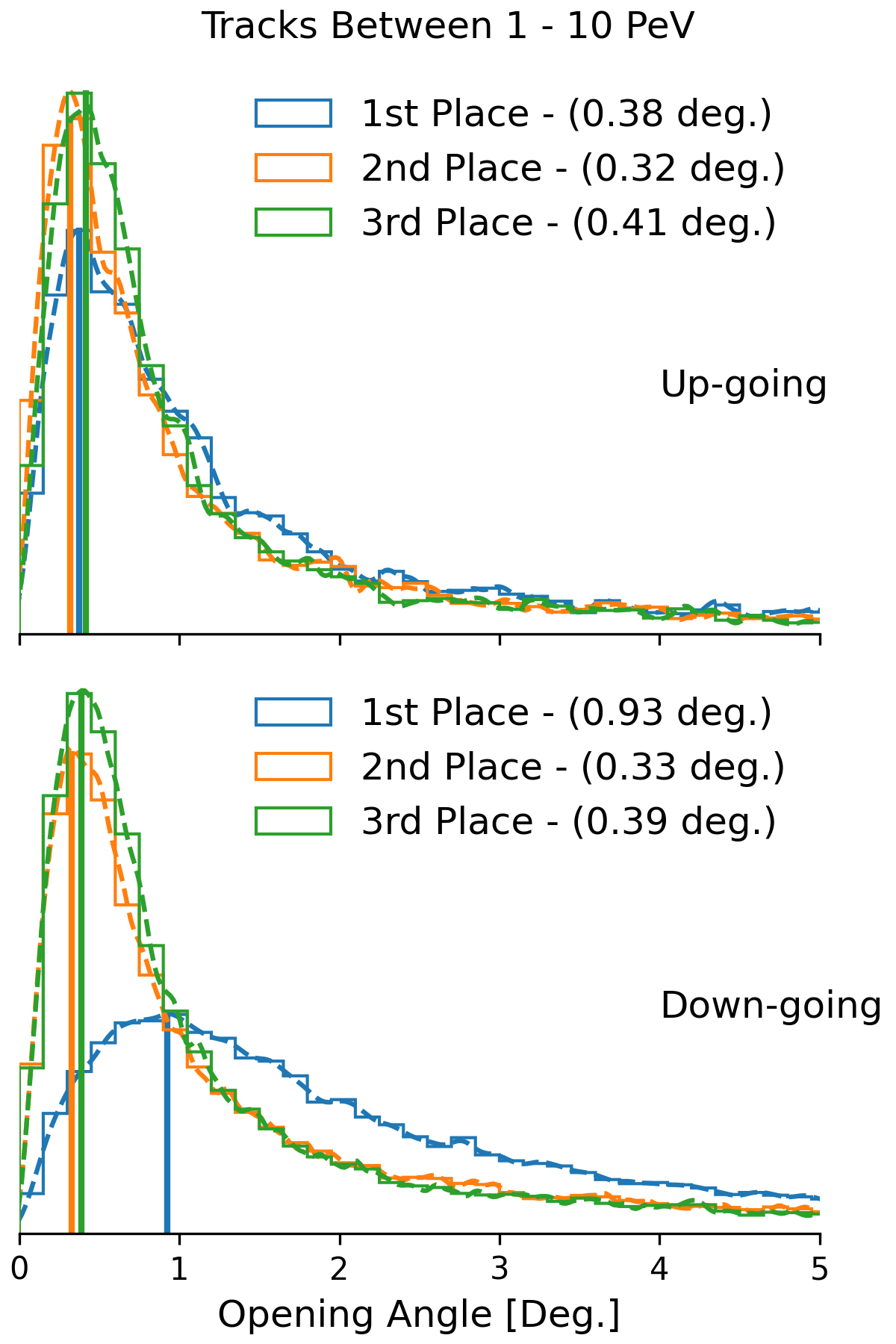}
    \caption{Example of  KDEs constructed for each solution on track events between 1 and 10 PeV. Estimated primary mode is denoted with the vertical lines. \textbf{Top:} Up-going track events. \textbf{Bottom:} Down-going track events.}
    \label{fig:kde_tracks}
    \end{center}
\end{figure}

The three solutions were evaluated on an similar but independent sample of nearly 1 million neutrino events of all flavours and interaction types. It originates from the same simulation used for the competition, but is sub-sampled such that it includes more events at the lower and higher energy range to facilitate comparisons there. 
Since this event sample---just like the training data in the competition---contains a lot of non-reconstructable events that consists of only noise pulses or atmospheric muons, the distribution of opening angles between reconstructed and true directions are relatively wide (see Figs.~\ref{fig:kde_cascades} and \ref{fig:kde_tracks}). In order to report numbers that correspond to the resolution of the well-reconstructable events, we decided to use the mode of the opening angle distribution instead of other summary statistics like the median or the mean which are strongly affected by the non-reconstructable events that span all opening angles between 0 and 180 degree almost uniformly distributed.
This procedure gives an approximation of the expected resolution on analysis level neutrino samples, where low quality events that occupy the wide tails have been excluded.

\begin{figure}[h!]
    \begin{center}
    \includegraphics[width=0.45\textwidth]{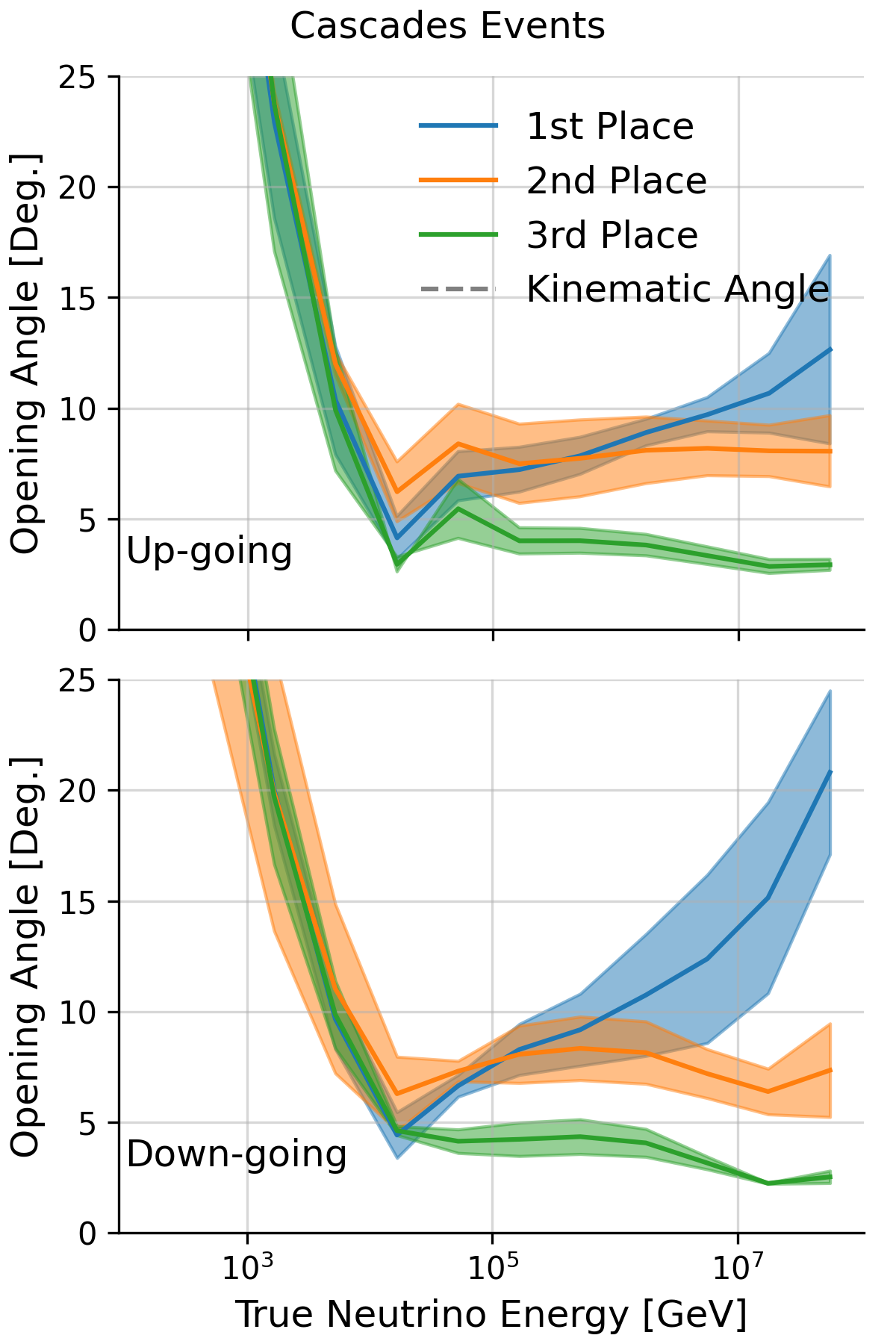}
    \caption{Estimated primary mode of opening angle vs. neutrino energy for cascade events. Each curve represents a solution. Solid lines depict estimated modes and bands denote 1 sigma uncertainty. \textbf{Top:} Estimated primary mode for up-going cascade events. \textbf{Bottom:} Estimated primary mode for down-going events}
    \label{fig:resolution_cascades}
    \end{center}
\end{figure}

\begin{figure}[h!]
    \begin{center}
    \includegraphics[width=0.45\textwidth]{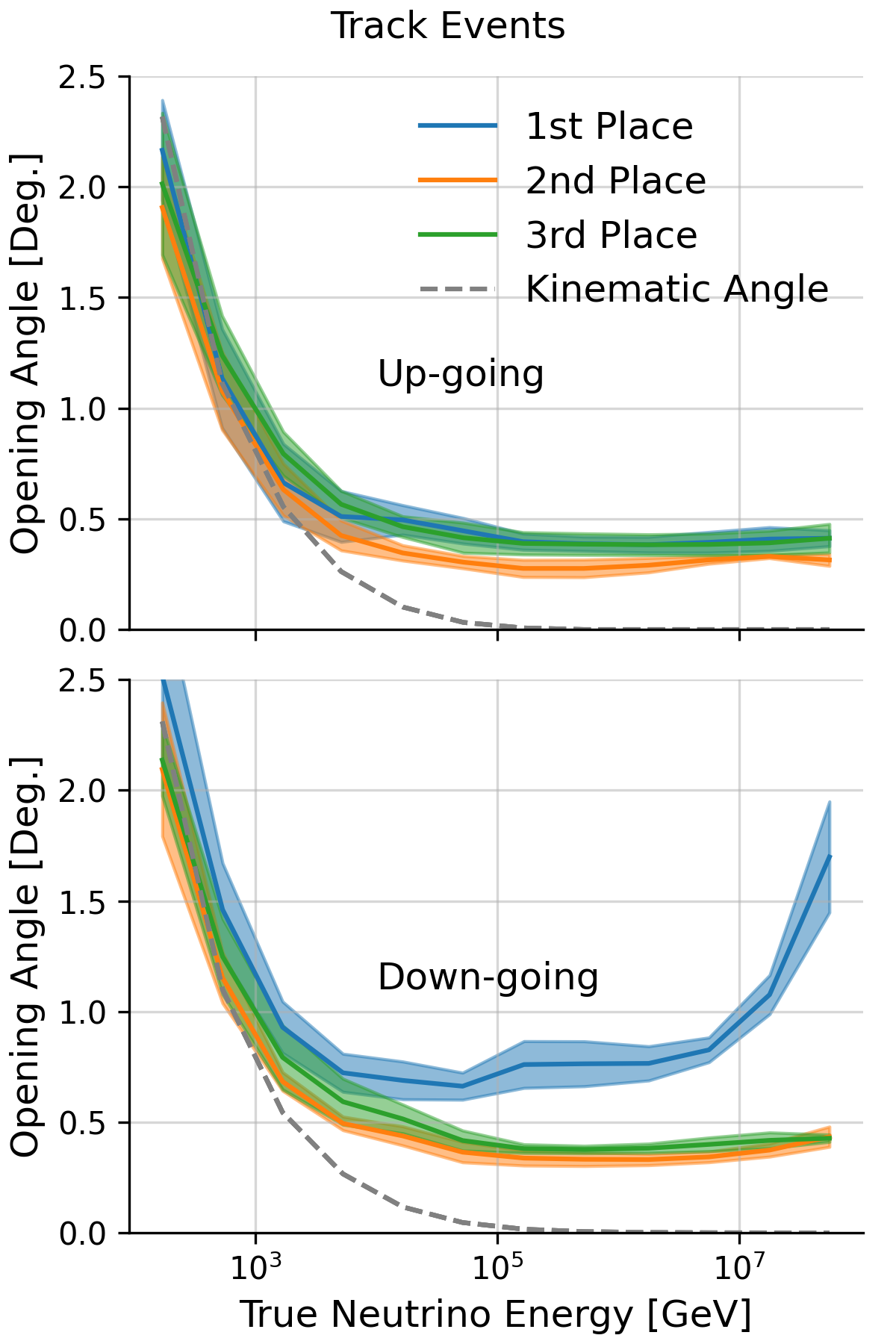}
    \caption{Estimated primary mode of opening angle vs. neutrino energy for track events. Each curve represents a solution, and the kinematic angle is shown in grey.}
    \label{fig:resolution_tracks}
    \end{center}
\end{figure}

The opening angle distribution is binned in true neutrino energy and for each energy bin, a \textit{kernel density estimator} (KDE) is used to estimate the primary mode of the distribution. Examples of such KDEs are also shown in Figs.~\ref{fig:kde_cascades} and \ref{fig:kde_tracks}. 
The comparisons of the achieved resolutions as a function of neutrino energy are provided for up- and down-going events separately, and are shown for cascade events in Fig.~\ref{fig:resolution_cascades} and tracks in Fig.~\ref{fig:resolution_tracks}.

As seen in Figure \ref{fig:resolution_cascades}, all methods have difficulties reconstructing cascade events up to energies of a few TeV, which is likely due to the composition of the competition dataset that provides only few events in this energy range, and that the signal-to-noise ratio becomes increasingly unfavourable as energy decreases. Beyond a few TeV, the \nth{1} place solution achieves its best of around 5 degrees for up- and down-going cascade events between 10 to 100 TeV but worsens as the energy increases. Similarly, the \nth{2} place solution reaches around 7 degrees for both regions, but stays below 10 degree towards the higher energy end. The \nth{3} place solution performs best across the energy range and reaches an opening angle of less than 5 degrees for the events with highest energies for both up- and down-going events. 

In Figure \ref{fig:resolution_tracks} the resolutions on track events is shown. An additional curve has been added that depicts the median kinematic opening angle between the initial neutrino and the outgoing muon from the interaction, representing the expected information limit from only reconstructing the track direction.
Around 1 TeV and lower, all methods follow the kinematic angle closely, but flatten beyond a few TeV for both up- and down-going tracks. The \nth{1} place solution settles below half a degree for up-going track events after around 10 TeV but for down-going tracks, it reaches a low of around 0.6 degrees at 10 TeV where after the resolution worsens considerably. The \nth{2} place solution settles under 0.5 degrees after around 5 TeV and reaches a low of around 0.4 degrees for both up- and down-going events. Although the \nth{3} place solution performs similarly to the \nth{1} place solution on the up-going tracks, it is closer to the \nth{2} place solution on the down-going tracks.  

From Figures \ref{fig:resolution_cascades} and \ref{fig:resolution_tracks} it is evident that the three solutions find different ways of minimizing the average opening angle: The \nth{1} place solution appears to resolve up-going tracks significantly better than down-going, the \nth{2} place solution reconstructs tracks very well, whereas the \nth{3} place solution appears to resolve cascade events significantly better than the other methods.

The  \nth{1},  \nth{2} and  \nth{3} solutions used sequence lengths of up to 6000, 768 and 3072 for inference with their transformers, respectively. These choices in hyper-parameters were conditioned on the competition dataset, where less than 2\% of the events had a sequence length larger than 800, making the number of events subject to significant sub-sampling small. Because the sequence length increases with neutrino energy, these choices in sequence length might be sub-optimal for the high-energy neutrinos shown in Figures \ref{fig:resolution_cascades}, \ref{fig:resolution_tracks} and further improvements in the high energy range might be possible.

%% file: sections/conclusion.tex
While thousands of solutions were submitted during the three-month ``IceCube -- Neutrinos in Deep Ice'' Kaggle competition, here we focused on the three winning ones.
Data processing, model architectures, and training procedures for each of the top three solutions were described in this article, providing insight into these novel ways of applying machine learning to IceCube data. While the three approaches have several aspects in common, they also differ in many ways. The final kaggle scores of the three solutions were almost on par, but differences can be observed when analyzing and comparing the performance in greater detail. In this work, a comparison differentiating between track and cascade events, as well as up- and down-going events, and shown as a function of the neutrino energy was provided. These comparisons reveal, although all three solutions provided highly accurate reconstructions in general, that there are stark differences. Overall, the \nth{3} place solution delivers the best cascade reconstruction with a resolution consistently below 5 degrees for neutrino events with energies above 10\,TeV. For the track events, the \nth{2} place solution performs overall best with resolutions that are consistently below the 0.5 degree mark for neutrino events with energies above 10\,TeV. 

The current IceCube state-of-the-art reconstruction that specifically targets cascade events reports a median angular resolution of 5-15 degree for a comparable energy range as quoted above \cite{icecube_dnn}. Another recently published IceCube reconstruction optimized for track events quotes median opening angles of around 0.2-0.6 degrees \cite{IceCube:2021oqo} in a similar energy range.
Although these numbers cannot be compared one-to-one with those reported by our comparisons, they indicate that the solutions found in the Kaggle competition are strong contenders and possibly able to outperform existing IceCube approaches, in particular for cascade events. 

It is also noteworthy that in the top 20 best performing solutions, the vast majority relied on a combination of graph convolutional neural networks and transformers. This provides a strong hint on how future ML-based reconstruction algorithms could look like in IceCube. 

Another strong suit of these machine learning-based Kaggle solutions is their high inference speed and their applicability to different event types, including noise and other backgrounds. This means that the methods can, in principle, be applied to the online IceCube data stream and provide unprecedented real-time reconstruction quality.

While further studies are needed from the IceCube collaboration that directly compare to existing IceCube algorithms on an equal footing, we were able to demonstrate the performance of the competition winners on the Kaggle dataset. Our resolution figures show an exciting potential of these new methods.